\documentclass[a4paper,10pt,aps,amsmath,amssymb, prapplied, twocolumn,superscriptaddress]{revtex4-2}

\usepackage{amssymb}
\usepackage{verbatim}

\usepackage{graphicx}
\usepackage{mathtools}
\usepackage{empheq}
\usepackage{enumerate}

\usepackage{dcolumn}
\usepackage{tikz}
\usepackage{bm}
\usepackage{bbm}
\usepackage[normalem]{ulem} 
\usepackage{multirow}

\usepackage[capitalise]{cleveref} 

\newcommand{\ket}[1]{|#1\rangle}
\newcommand{\bra}[1]{\langle#1|}

\newcommand{\Tr}[1]{\text{Tr}\left[#1\right] }
\newcommand{\prom}[1]{\langle #1\rangle }
\newcommand{\1}[0]{\mathbbm{1}}

\begin{document}

\title{Closest Accessible Symmetry reduction: a tool for Hamiltonian interpolation analysis}

\author{Ana Palacios}
\email{ana.palacios@qilimanjaro.tech}
\affiliation{%
 Qilimanjaro Quantum Tech, Carrer de Veneçuela, 74, Sant Martí, 08019, Barcelona, Spain}
\affiliation{Departament de F\'{i}sica Qu\`{a}ntica i Astrof\'{i}sica, Facultat de F\'{i}sica,
Universitat de Barcelona, 08028 Barcelona, Spain}
\affiliation{Institut de Ci\`{e}ncies del Cosmos, Universitat de Barcelona,
ICCUB, Mart\'{i} i Franqu\`{e}s 1, 08028 Barcelona, Spain.}
\author{Artur Garcia-Saez}
\affiliation{%
Qilimanjaro Quantum Tech, Carrer de Veneçuela, 74, Sant Martí, 08019, Barcelona, Spain}
\affiliation{Barcelona Supercomputing Center, Plaça d'Eusebi Güell, 1-3, Les Corts, 08034 Barcelona, Spain}
\author{Arnau Riera}
\affiliation{Qilimanjaro Quantum Tech, Carrer de Veneçuela, 74, Sant Martí, 08019, Barcelona, Spain}

\author{Marta P. Estarellas}%
\affiliation{%
Qilimanjaro Quantum Tech, Carrer de Veneçuela, 74, Sant Martí, 08019, Barcelona, Spain}

\date{\today}

\begin{abstract}
We introduce a framework for analysing the spectrum of Hamiltonian interpolations without heavily relying on discretising the interpolation parameter. The method is based on the concept of accessible symmetries: a problem-class-dependent family of certifiable reflections that induce bipartitions of Hilbert space. At each step, the interpolation Hamiltonian is projected onto the sectors of the accessible symmetry that is closest to being satisfied, yielding a hierarchy of weakly coupled pseudo-eigenspaces together with explicit residual couplings between them. We show that this representation captures qualitative signatures of quantum phase transitions, provides estimates of their location, and offers insights into their nature. The quality of the approximation is controlled by the compatibility between the accessible symmetry family and the problem instance. Although motivated by adiabatic quantum computation, our approach applies more broadly to the study of Hamiltonian phase diagrams, providing a new perspective on the spectral reorganisation of many-body quantum systems.
\end{abstract}

\maketitle


\section{Introduction} 

Spectral information is extremely valuable for the analysis of the interpolation process between two Hamiltonians. In fact, in many cases it is precisely the insight we are after: the location of a phase transition, set where the energy gap between ground and first excited states closes; or the efficiency of an adiabatic quantum computation (AQC) algorithm, set by the size of said energy gap along the computation in accordance with the adiabatic theorem~\cite{born_beweis_1928,kato_adiabatic_1950,jansen_bounds_2007,comparat_general_2009,amin_consistency_2009,bachmann_adiabatic_2017,albash_adiabatic_2018}.

Currently, there is a variety of different methodologies to approach the study of a Hamiltonian interpolation, such as tensor networks~\cite{white_dmrg_1993,verstraete_dmrg_2004,paeckel_tebd_2019,cirac_mpsreview_2021,wei_efficient_2023}, geometric approaches~\cite{zanardi_ground_2006,zanardi_information-theoretic_2007,rezakhani_intrinsic_2010,kumar_geodesics_2012}, exact diagonalisation (e.g. Lanczos) or semi-definite programming relaxations~\cite{jansen_mapping_2026} . 
Despite their differences, these approaches typically infer the properties of the interpolation from information obtained at individual points or local regions of parameter space, often studying the instantaneous objects along the full interpolation.
%
We argue that there exists a form of conceptual imbalance in these programs: all efforts are poured into unveiling the structure of intermediate Hamiltonians when, at least in the context of a convex interpolation, all the information is already determined by the structures of the initial Hamiltonian, $A$, and the final one, $B$. 
This raises the natural question of whether useful spectral information about the entire interpolation can be extracted from the joint structure of $A$ and $B$ alone.
In this work, we answer this question affirmatively through a recursive approximate joint symmetry analysis of both Hamiltonians. The resulting construction yields a single hierarchical description of the interpolation rather than a collection of local descriptions at different values of the interpolation parameter.

The mention of approximate joint symmetry analysis is indeed resonant with the Schrieffer-Wolff (SW) transformation~\cite{schrieffer_SW_1966,bravyi_SW_2011}, which is typically employed for understanding a perturbed version of some well-understood Hamiltonian. However, the regime of applicability of SW is limited to the existence of a direct rotation between sectors of Hilbert space of equal dimension, and it is only useful whenever such direct rotation connects the low-energy sector of the unperturbed Hamiltonian to the low-energy sector of the perturbed one. For this reason, SW ceases to provide a globally valid description whenever the interpolation to the perturbed Hamiltonian crosses over a critical region, since the spectral gap between the sectors that we are trying to decouple will generally close in this scenario. In contrast, our method forces the splitting of the Hilbert space even when this induces a large error, which allows us to treat arbitrary pairs of Hamiltonians. Since the error is still minimised within certain constraints, as well as kept track of, these potentially large errors do not necessarily undermine the effectiveness of the method. All in all, unlike SW constructions, which derive effective descriptions around a chosen reference Hamiltonian, our method is fundamentally linked to a chosen pair of reference Hamiltonians, where both are treated on equal footing.

More accurately, our approach is similar in spirit to multiscale matrix factorisation methods~\cite{pmlr-v32-kondor14} but adapted to the Hamiltonian interpolation setting, where we need to consider both simultaneous block diagonalisation and the efficient representation of quantum systems. Importantly, instead of searching over the full unitary space for block-diagonalising transformations, we restrict the search to a reduced family of closest accessible symmetries (CAS). These correspond to tractably certifiable bipartitions of Hilbert space that are physically motivated by the structure of the problem at hand. Then, rather than searching for exactly preserved subspaces, we recursively identify the most weakly coupled bipartitions of Hilbert space that are accessible within this tractable family of certifiable symmetries, and subsequently project onto their associated symmetry sectors. In particular, this CAS reduction can be iterated until only $2\times2$ matrices remain, which we can diagonalise exactly as a function of an interpolation parameter $s$.
This provides us with a hierarchical decomposition that progressively isolates an approximation to the low-energy structure of the problem by iteratively truncating the couplings between weakly interacting sectors of the spectrum. 
This decomposition yields an approximate pseudo-eigenspectrum together with explicit off-diagonal couplings that mediate hybridisation between pseudo-eigenspaces. The combination of these provides valuable insights about the true spectrum, since the pseudo-gap structure plus the relevant off-diagonal couplings enable the (approximate) identification of perturbative anticrossings, which are important first order phase transitions in the context of AQC, and, more generally, the diagnosis of different kinds of quantum phase transitions. We highlight that the resulting pseudo-spectrum constitutes a reduced representation of the true interpolation; a compression whose quality depends on the compatibility between the accessible symmetry family and the problem instance at hand.

Given this picture, one can note that our approach also bears some conceptual resemblance with renormalisation group (RG) methods. More specifically, it is reminiscent of real-space RG approaches~\cite{cardy_scaling_1996,fisher_disorderedRG_1995}, where reduced descriptions emerge through successive reductions to effective degrees of freedom. However, unlike conventional RG transformations, the CAS reduction is not defined through scale transformations or assumptions of self-similarity. In fact, the effective Hamiltonians of different branches of the recursion are generally inequivalent and need not preserve the formal structure of the original problem. This distinction sets apart our approach from the corpus of RG methods.

The structure of the paper is as follows: In Section~\ref{sec:CAS_definition} we introduce the concept of accessible symmetries and describe the procedure to build accessible symmetry sets. Section~\ref{sec:analytical_approx_of_spectrum} presents the hierarchical decomposition induced by the CAS reduction and the information that can be extracted from it about the true interpolation spectrum. All this theory is given a concrete shape in Section~\ref{sec:TFIM_example}, where we define a CAS set for interpolations between transverse-field Hamiltonians and arbitrary Ising models, which constitutes the standard quantum annealing setting. Numerical results regarding this concrete example are presented in Section~\ref{sec:numerical_results}, and we discuss the conclusions and outlook of this work in Section~\ref{sec:conclusions}.

\section{Closest Accessible Symmetry (CAS)}\label{sec:CAS_definition}

Before jumping straight into Hamiltonian interpolations, we first introduce the core concept of this work by considering a single arbitrary, $d$-dimensional Hamiltonian $H$, where $d=2^N$ is the full Hilbert space dimension. 
Let us say we are interested in finding a symmetry $T$ of $H= \sum_k E_k \ket{E_k} \bra{E_k}$, such that $[T, H] = 0$, which splits the Hilbert space in half. Without loss of generality, it is convenient to define $T$ as a reflection, such that $T = 2P_0 - \1 $ with $\Tr{P_0} = d/2$. We can then define this search as the following optimisation problem:
\begin{gather}
    \min_{P_0} || P_0 H (\1-P_0) + (\1-P_0) H P_0||_2 
    \label{eq:fun2minimise_singleH} \\
    \text{s.t. } 
    \left\{
    \begin{alignedat}{3} 
        & P_0^2 = P_0, \\
        & \Tr{P_0} = \frac{d}{2}
    \end{alignedat}
    \right.
    \label{eq:conditions_fun2minimise_singleH}
\end{gather}
where $||H||_2$ denotes the operator norm of $H$, $||H||_2 = \sup_{||\vec{x}||=1} ||H\vec{x}||$. For nondegenerate $H$, it can be seen that the optimal solutions of~\cref{eq:fun2minimise_singleH,eq:conditions_fun2minimise_singleH} are all the $\begin{pmatrix}
    d \\ d/2
\end{pmatrix}$ sums of possible combinations of spectral projectors $\ket{E_k}\bra{E_k}$, e.g. $P_0 = \sum_{i=0}^{d/2 - 1}\ket{E_k}\bra{E_k}$ and $P_1 = \sum_{i=d/2}^{d-1}\ket{E_k}\bra{E_k}$ for $E_0 < E_1 < ... < E_{d-1}$. The space of optimal solutions is even larger if $H$ contains some degeneracies, since now there are some eigendirections that are not uniquely defined. Once we fix one such $P_0$, we can consider the projections $H_0 = P_0 H P_0$, $H_1 = (\1 - P_0) H (\1 - P_0)$ and repeat the same process for $H_0, H_1$ and $d' = d/2$. Iterating this process $N-1$ times and solving the problem on the $2^{N-r}$ blocks at each step will provide us with the full diagonalisation of the system, and thus performing this process exactly is just as hard as finding the full spectral decomposition of $H$. 

So far, this just looks like a more convoluted way of formulating the problem of spectral decomposition than usual. The optimisation of Eq.~\eqref{eq:fun2minimise_singleH} takes place over a highly nonconvex manifold~\eqref{eq:conditions_fun2minimise_singleH}, and even the first iteration will be generally as hard as finding the spectral decomposition of $H$, since finding half of it is not significantly easier than finding it in full. However, while equally hard to solve exactly, we can construct an easier program that allows to approximately solve~\cref{eq:fun2minimise_singleH,eq:conditions_fun2minimise_singleH} by restricting the search to a family of $T$s that we know have the correct shape and which can be argued to not be too far from commuting with $H$. For ensuring the right shape, by construction we only consider families of reflections acting on half of the Hilbert space. The argument is that minimising the commutator with $H$ within this family can be done efficiently if $H$ belongs to a sufficiently restricted class of Hamiltonians, as will be exemplified in Section~\ref{sec:TFIM_example}.
Thus, we relax our problem to the minimisation
\begin{equation}
    \min_{P_0 \in \mathcal{A}} || P_0 H (\1-P_0) + (\1-P_0) H P_0||_2 \ , 
    \label{eq:fun2minimise_singleH_accessible}
\end{equation}
where $\mathcal{A}$ is the family of $d/2$-projectors associated to the family of symmetries $T$ we are able to certify, $\mathcal{T} =\{T=2P-\1 | P \in \mathcal{A}\}$, which we then consider accessible. Thus arises the concept of Closest Accessible Symmetry (CAS): the bisection of the Hilbert space that produces the most invariant subspaces within a given family of proper bisections.
Note that the restriction to $\mathcal{A}$ removes the need for the constraints in Eq.~\eqref{eq:conditions_fun2minimise_singleH}, as they are now incorporated in the definition of $\mathcal{A}$. Let us define the cost
\begin{equation}
	\varepsilon(\mathcal{A}) = \min_{P \in \mathcal{A}} || P H (\1-P) + (\1-P) H P||_2 \ .
	\label{eq:error_accessible_set}
\end{equation}
Then, by considering increasingly general families of $\mathcal{S}_k$ as accessible, we can build the following hierarchy:
\begin{gather}
	\mathcal{A}_0 \subseteq \mathcal{A}_1 \subseteq ... \subseteq \mathcal{A}_{\text{full}} \ ,  \\
	\varepsilon(\mathcal{A}_0) \geq \varepsilon(\mathcal{A}_1) \geq ... \geq \varepsilon(\mathcal{A}_{\text{full}}) = 0 \ ,
\end{gather}
where $\mathcal{A}_{\text{full}}$ corresponds to the set containing all projectors satisfying~\eqref{eq:conditions_fun2minimise_singleH}. We can see this hierarchy as the interpolation between restricted but tractable searches and the exact (but intractable) symmetry search.

The quality of the approximation will depend on the closeness of our accessible set to the exact symmetries of the problem instance in question. Thus, in order to obtain information about the low-energy spectrum of a single Hamiltonian, standard approaches such as Lanczos will be more cost-efficient. However, the CAS approach is interesting in the context of examining the low-energy spectrum along an interpolation because it splits weakly coupled subspaces throughout the evolution as a function of the interpolation parameter $s$, thus avoiding the discretisation of time and potentially becoming more cost-efficient than other approximation techniques that need to rely on it. Thus, in the following we'll focus on the case of the time-dependent $H$ we are concerned with in an interpolation. 
For simplicity, we'll consider the standard linear interpolation
\begin{equation}
    H(s) = (1-s) A+ s B
    \label{eq:linear_interp_H}
\end{equation}
for 2-local, spin Hamiltonians $A$ and $B$ and $s\in [0, 1]$.
Note that the gap profile of more general schedules, such as $H(s) = \lambda_A(s) A + \lambda_B(s)B$ with $\lambda_A(0) = \lambda_B(1) = 1$, $\lambda_A(1) = \lambda_B(0) = 0$, is related to that of the linear interpolation as
\begin{equation}
    H = (\lambda_A + \lambda_B) \left[ \left( 1 - \frac{\lambda_B}{\lambda_A + \lambda_B} \right) A + \frac{\lambda_B}{\lambda_A + \lambda_B} B \right]
\end{equation}
as long as $\lambda_A + \lambda_B > 0$.
Without loss of generality, we'll additionally consider $A$ and $B$ to be traceless and non-commuting.

\subsection{Building the accessible set}\label{sec:accessible_set}

We hereby describe the procedure to build restricted, accessible sets for an interpolation Hamiltonian as the one in~\eqref{eq:linear_interp_H} such that the 
direct optimisation of~\cref{eq:fun2minimise_singleH,eq:conditions_fun2minimise_singleH}, which is highly complex, can be spared.

The key step forward consists on removing the constraints from~\eqref{eq:conditions_fun2minimise_singleH}, which means that we need to ensure that we are always searching through $\mathbb{Z}_2$-type symmetries, and we find the key tool to obtain these $\mathbb{Z}_2$-symmetry certificates in dynamical Lie algebra (DLA) analysis.
Our goal is not to compute the full DLA, but to efficiently certify whether it has the block structure of two invariant sectors of equal dimension under a proposed modification of the Hamiltonians, which in our case will consist of term removal.

The DLA of a given Pauli set $\Omega$ is the Lie closure of its elements, that is, the algebra generated via nested commutators of the elements in $\Omega$, more explicitly
\begin{align}
    &\text{DLA}(\Omega) = \nonumber\\
    &=\text{span}_{\mathbb{R}}\{X_{i_0}, [X_{i_1}, ..., [X_{i_{k-1}}, X_{i_k}] ...] | X_{i_j} \in \Omega , k \geq 1\} \ .
\end{align}
In the context of Hamiltonian dynamics, the DLA characterizes the set of transformations that can be generated by a system and therefore plays a central role in the theory of quantum control~\cite{dalessandro_introduction_2021}. More recently, DLAs have also become an important tool in quantum computing, where they have been used to characterize the expressivity and trainability of parametrized quantum circuits~\cite{larocca_diagnosing_2022,fontana_characterizing_2024,ragone_a_2024} and study the classical simulability of quantum circuits through the framework of $\mathfrak{g}$-sim~\cite{goh_lie_2025,barligea_enabling_2026}.

We consider the identification of the CAS set $\mathcal{T}$ by analysing the DLA generated by the alphabet of Pauli terms $\Omega_{A \cup B}=\Omega_A \cup \Omega_B$ present in $H(s)$. Let us describe an arbitrary Hamiltonian in the Pauli basis as
\begin{equation}
    X = \sum_i c^X_i \hat{\sigma}_i\ ,
\end{equation}
where $\hat{\sigma}_k$ refers to the Pauli string we have indexed as the $k$-th basis element of Pauli space. Then, its associated alphabet is
\begin{equation}
    \Omega_{X} = \{\hat{\sigma}_i | \Tr{X \hat{\sigma}_i}\neq 0 \} \ .
\end{equation}
In turn, the alphabet we are concerned with for the CAS set definition is
\begin{gather}
    \Omega_{A \cup B} = \{\hat{\sigma}_i | \Tr{A \hat{\sigma}_i}\neq 0 \lor \Tr{B \hat{\sigma}_i}\neq 0 \} \ .
\end{gather}
As a concrete example, let us consider $A= \frac{1}{\sqrt{2}}(X_0 + X_1)$ and $B= 0.76Z_0 + 0.4X_0 - 0.1 Z_1 + 0.5 Z_0 Z_1$. Then, we have $\Omega_A = \{X_0, X_1\}, \Omega_B = \{X_0, Z_0, Z_1, Z_0Z_1\}$ and $\Omega_{A \cup B} = \{ X_0, X_1, Z_0, Z_1, Z_0Z_1 \}$.

For the determination of the CAS set, we will use the fact that $\text{DLA}(\Omega_{A \cup B})$ has a block-diagonal structure in the presence of a shared $\mathbb{Z}_2$ symmetry, which can be represented by a single Pauli string in some basis. That is, when we have $T=\hat{\sigma}_{\alpha}$ such that $[A, T] = [B, T] = 0$, we also find $\text{DLA}(\Omega_{A \cup B}) = \mathfrak{su}(d/2) \oplus \mathfrak{su}(d/2)$. 

Notice that the restriction to single-Pauli string $T$'s is a consequence of using bare Pauli strings as the generators of the DLA instead of linear combinations of them taking into account the $c^A_i, c^B_i$. This is the core restriction we are imposing that limits the expressivity of our CAS set $\mathcal{T}$ (and the associated set of accessible projectors $\mathcal{A}$). Due to the independent treatment of Pauli strings, the CAS search effectively amounts to identifying the minimal set of terms that must be removed from $A$, $B$ (with respect to the cost~\eqref{eq:error_accessible_set}) such that we reach the desired structure of $\text{DLA}(\Omega_{A \cup B})$. Let us now define
\begin{align}
	\tilde{A} &= A - M_A \ ,\\
	\tilde{B} &= B - M_B \ ,
\end{align}
where we have identified a CAS operator $T$ such that $[T, \tilde{H} ] = 0$ with $\tilde{H} = (1-s)\tilde{A} + s\tilde{B}$ and $M_A, M_B$ are the parts of $A$, $B$ associated with the terms in $\Omega_{A \cup B} \setminus \Omega_{\tilde{A} \cup \tilde{B}}$. Thanks to the understanding of DLA structure of Pauli sets that the community has gained in the recent years~\cite{wiersema_classification_2024,kokcu_classification_2024,aguilar_full_2024}, a systematic analysis is possible for sufficiently restricted families of Hamiltonians, as exemplified in Section~\ref{sec:TFIM_example}. Efficient heuristic algorithms could be built for more complex Hamiltonian families, e.g. extending the methods of~\cite{aguilar_full_2024}. If the bases of $A$ and $B$ are not orthogonal from the start, better accessible symmetries can be achieved by running a prior optimisation to search for a basis in which the description of both $A' = U^\dagger A U$ and $B' = U^\dagger B U$ is maximally compact in Pauli space (i.e., minimising the support of $A', B'$ simultaneously). This is due to the fact that, as previously stated, we only consider access to single-Pauli string symmetries $T'=\hat{\sigma}_{\alpha}$. As described in Appendix~\ref{sec:compression_pauli_space}, this compression process can be readily tackled via gradient flow methods.

\section{Analytical approximation to the interpolation spectrum}\label{sec:analytical_approx_of_spectrum}

Now that we are acquainted with the notion of CAS, in the present section we assume that we have defined a CAS set that is consistent with our ability to search through it efficiently.
Equipped with said CAS search strategy, we now apply it recursively to the blocks we split in $A$ and $B$ until we end up with $2\times 2$ matrices, which we can then diagonalise exactly.

\begin{figure}
    \centering
    \includegraphics[width=\linewidth]{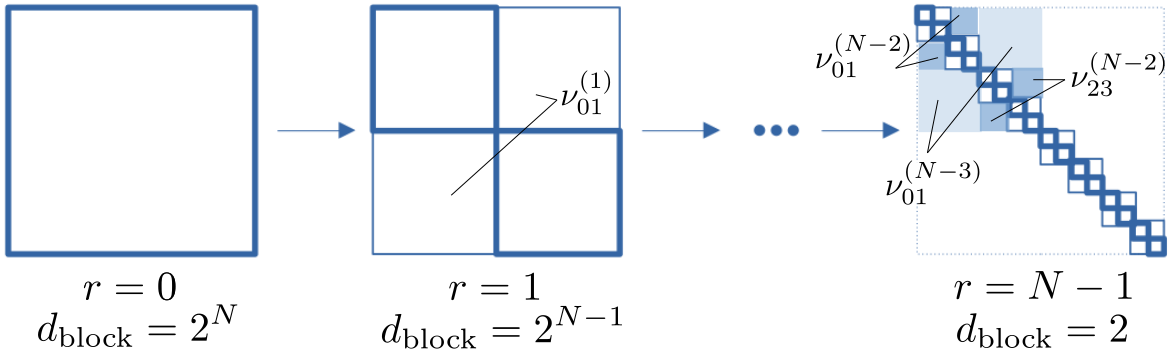}
    \caption{Illustration of the consecutive block reduction process of $H^{(r)}$, where in each iteration $r$ every block is split in half. Consequently, every iteration a set of $\{P_{0, (\eta)}^{(r)}, M_{A, (\eta)}^{(r)}, M_{B, (\eta)}^{(r)}\}_{\eta=0}^{2^{r} -1}$ is found. 
    }
    \label{fig:consecutive_block_reduction}
\end{figure}

We will refer to the CAS obtained in an iteration at recursion depth $r$ as a level $r$ CAS. If fully parallelised, the $2\times2$ matrices can be reached in $N$ iterations, where for each iteration we have $2^r$ blocks, out of which we extract a level $r$ CAS of dimension $2^{N-r}$ (see overview in Fig.~\ref{fig:consecutive_block_reduction}).
However, due to the exponential scaling of the number of branches to keep track of as $r$ increases, the computation of every single one of them is not realistic even with parallelisation strategies.
Nonetheless, provided that we can identify low-energy branches, tracking a reduced set is enough to obtain the information relative to the low-energy spectrum.

The aforementioned procedure allows us to split the full Hamiltonian in a set of pseudo-eigenvalues $\{\mu_k\}$ and pseudo-eigenvectors  $\{\ket{\mu_k}\}$ plus the off-diagonal parts $\nu$ that have been separated at each step of the process. At recursion order $r$ we have the block-diagonals produced at that iteration, $H^{(r)}_{\eta_r}$, the associated block off-diagonals, $\nu^{(r)}_{2\eta_r, 2\eta_r + 1}$ and the block off-diagonals that were split in previous iterations of order $q<r$, $\nu^{(q)}_{2\xi_q, 2\xi_q + 1}$, as shown in the right-most part of Fig~\ref{fig:consecutive_block_reduction}. Thus, the description of $H^{(r)}$ becomes
\begin{equation}
    H^{(r)} = \sum_{\eta_r=0}^{2^{r-1}} \left[ H^{(r)}_{\eta_r} + \nu^{(r)}_{2\eta_r, 2\eta_r + 1} \right] + \sum_{q = 0}^{r-1} \sum_{\xi_q=0}^{2^{q-1}} \nu^{(q)}_{2\xi_q, 2\xi_q + 1} \ .
\end{equation}
At the final recursion step, $r=N-1$, we find the pseudo-eigenvalues $\mu_k=\mu_{\eta_r}^{\pm}$ and pseudo-eigenspaces $\Pi_{\eta_r}^{\pm}=\ket{\mu_k}\bra{\mu_k}$ that we take as a raw approximation to the true eigenspectrum.
\begin{equation}
    H^{(N-1)}_{\eta_r} = \mu_{\eta_r}^{+} \Pi_{\eta_r}^{+} + \mu_{\eta_r}^{-} \Pi_{\eta_r}^{-} \quad \eta =0, ..., 2^{N-1}
\end{equation}
It is helpful to represent the indices of the emerging blocks as ``paths", recording whether we projected to the +1 or -1 sector of the CAS at each recursion order $r$. Note that, at any order $r$, the interaction between the subspaces of the split blocks is contained in the off-diagonal $\nu$ connecting them; no other prior off-diagonal block ($r'<r$) participates. This is true independently of the size of $||\nu||$. Thus, given the distinctive paths that label two pseudo-levels, $\vec{p}, \vec{q}$, the off-diagonal block  that connects them $\nu_{\vec{p}}^{\vec{q}}$ will be the one corresponding to the split where both paths diverge, such that $\nu_{\vec{p}}^{\vec{q}} = \nu_{\vec{\xi}}$ for $\vec{p} = (\vec{\xi}, \vec{\xi}_0)$, $\vec{q} = (\vec{\xi}, \vec{\xi}_1)$ with $[\vec{\xi}_0]_0 \neq [\vec{\xi}_1]_0$. In this manner, as long as there are no near degeneracies with other pseudo-levels (see upcoming section), their interaction will be bounded by the spectral norm of the connecting off-diagonal, $||\nu_{\vec{\xi}}||_2$.

At any point, we can identify the contributions to the off-diagonals as a part coming from $A$,
\begin{equation}
    M_A^{tot} = \sum_r \sum_{\eta_r} M_{A, (\eta_r)}^{(r)}(s) \ ,
\end{equation}
and a part coming from $B$, 
\begin{equation}
    M_B^{tot} = \sum_r \sum_{\eta_r} M_{B, (\eta_r)}^{(r)}(s) \ .
\end{equation}
We can thus build the full off-diagonal $\nu$ from the consecutive $M_A, M_B$ as
\begin{gather}
    \nu^{tot}(s) = \nu^A + \nu^B = (1-s) M_A^{tot} + sM_B^{tot} \ ,\\
    H(s) = \sum_k \mu_k(s)  \Pi_k(s) + \nu^{tot}(s) \ .   
\end{gather}
Note that, as explicitly stated in the above expressions, $M_A$ and $M_B$ may be $s$-dependent themselves (due to an $s$-dependence of some CAS at some level of recursion). Thus, we need to establish a criterion to choose the optimal CAS despite them having different $s$-dependencies. In our numerical experiments we consider the cost
\begin{equation}
    \varepsilon= \mathcal{C}_{\max}(\nu) = \max_s ||\nu(s)||_2 \ .
    \label{eq:nu_max_cost} 
\end{equation}
Another possibility would be minimising the average of $||\nu(s)||_2$. We explored this alternative in our numerical experiments but, since it resulted in less accurate predictions, we relay these results to Appendix~\ref{sec:integral_cost}.
As we'll discuss in the concrete example of Section~\ref{sec:TFIM_example}, it's not always possible to efficiently obtain $||\nu(s)||_2$ (e.g. in the case in which there are non-commuting terms contributing to $\nu$), but we can always efficiently upper bound it by considering 
\begin{gather}
    \nu = \sum_i c_i \hat{\sigma}_i \ , \\
    ||\nu||_2 \leq \sum_i |c_i| = C  \ ,
    \label{eq:C_upper_bound_nu2}
\end{gather}
where once again $\hat{\sigma}_i$ denotes the $i$-th string of the Pauli basis.

We highlight that keeping track of the pseudo-eigenspaces $\{\Pi_k\}$ is not scalable, since the representation of a single pseudo-eigenstate will take exponential resources in the number of qubits. For this reason, in this work we will mostly restrict ourselves to the analysis of the pseudo-eigenspectrum $\tilde{\Lambda}$.

\subsection{Bounds on the spectral gap}\label{sec:bounds}

The pseudo-spectrum $\tilde{\Lambda} = \{\mu_{\xi}\}$ is our raw approximation of the true spectrum of $H(s)$, and from it we obtain a prediction $\tilde{\Delta}(s)$ of the instantaneous gap between ground and first excited states. This raw estimate is, of course, unreliable; for one, since the pseudo-levels $\mu_{\xi}$ are non-interacting (with the exception of their pair within the same $2\times2$ block), we will most often find that these pseudo-levels cross, which can never be a feature of the true spectrum in finite dimensions if $A$ and $B$ shared no exact symmetry.

To quantify when such crossings are physically meaningful and when they are simply a product of the poor approximation of our CAS reduction, we introduce the notion of hybridisation between pseudo-states. This concept captures when the disregarded interaction between pseudo-levels is relevant enough that we cannot treat them independently. To formalise this, let us define the hybridisation $\overline{\chi}$,
\begin{equation}
    \overline{\chi}_{kj} := \frac{|\bra{\mu_k} \nu \ket{\mu_j}|}{\tilde{\Delta}_{kj}} \ ,
    \label{eq:true_hybridisation}
\end{equation}
where $\tilde{\Delta}_{kj} = |\mu_k - \mu_j|$ is the gap between pseudo-levels $k$ and $j$. From here on, we will generally refer to $\tilde{\Delta}_{10}$ as the pseudo-gap.
Notice that $\overline{\chi}_{kj} \ll 1$ implies that those two pseudo-levels won't hybridise much, while larger $\overline{\chi}_{kj}$ values will imply increasingly higher hybridisation. Since accessing the off-diagonals $|\bra{\mu_k} \nu \ket{\mu_j}|$ is generally too expensive, we define an upper bound $\chi$ to the true hybridisation that we can efficiently keep track of,
\begin{equation}
    \chi_{kj} := \frac{||\nu_{kj}||_2}{\tilde{\Delta}_{kj}} \ ,
    \label{eq:hybridisation}
\end{equation}
where $\nu_{kj}$ is the off-diagonal block coupling the corresponding pseudo-eigenspaces and $||\nu||_2 = \lambda_{\max}(\sqrt{\nu^\dagger\nu})$ is its 2-norm. Unless explicitly stated otherwise, we will refer to this computable quantity $\chi_{kj}$ when discussing hybridisation.

We highlight that the hybridisation criterion remains sufficient only if the discarded $\nu$'s eigenbases are far from the true eigenvectors of the system, i.e., if the CAS reduction had a low score as per ~\eqref{eq:nu_max_cost}. If this is not the case, $\nu$ induces an important diagonal correction onto the pseudo-levels instead of an off-diagonal one, which can lead to reordering of pseudo-levels and the generation of low-lying, strongly hybridised clusters that would not be visible without taking this correction into account. Since this is an operational issue that depends on the compatibility between the instance under analysis and our particular CAS set, in this section we will not consider any pseudo-level reordering of such nature and instead take hybridisation as the only relevant factor.
Nonetheless, in Appendix~\ref{sec:restoration_bounds} we identify a scenario where this correction is needed within the specific Hamiltonian family under consideration in Section~\ref{sec:TFIM_example} and walk through the procedure to restore the correctness of the bounds. 
In the following, we will present the bounds on the spectral gap that apply for the different scenarios of low-lying pseudo-level hybridisation, relaying the detailed proofs to Appendix~\ref{sec:proof_bounds_gap}.
Whenever two pseudo-levels remain weakly hybridised with the rest, we can rigorously bound the incurred error on the gap $|\Delta(s) - \tilde{\Delta}(s)|$ making use of Weyl's inequalities~\cite{bhatia_matrix_1997}
. The latter can be combined with the Feshbach-Schur map~\cite{zhang_schur_2005,griesemer_smooth_2008} whenever we can be certain that both eigenvalue clusters won't overlap, which is indeed guaranteed for $||\nu_{kj}||_2 < \tilde{\Delta}_{kj} / 2$, thus providing a tighter bound whenever the two pseudo-eigenstates in question are far apart enough. Note that this condition is exactly equivalent to $\chi_{kj} < 1/2$; for this reason we will set $\chi^*_{kj} = 1/2$ as the numerical threshold to decide when hybridisation is important. However, we highlight that this threshold will be overly demanding in general, since $||\nu||_2$ is generally loosely bounding the relevant off-diagonal element in the pseudo-eigenbasis.
We hereby state the resulting bounds, 
\begin{gather}
    |\Delta_{kj} - \tilde{\Delta}_{kj}|   \leq
    \left\{
    \begin{array}{ll}
        \frac{2 \left(||\nu_{kj}||_2\right)^2}{\tilde{\Delta}_{kj} - ||\nu_{kj}||_2} & \text{if } ||\nu_{kj}||_2 < \tilde{\Delta}_{kj} / 2 \\
        2 ||\nu_{kj}||_2 & \text{otherwise}
    \end{array}
    \right.
    \label{eq:bounds_gap}
\end{gather}
where $\Delta_{kj} = E_k - E_j$ refers to the true gap between levels $k$ and $j$.

Whenever we encounter a cluster of strongly hybridised pseudo-levels, however, further considerations must be made and thus the expression of the bounds changes. 

We first introduce the collection of pseudo-level indices involved in the relevant strongly hybridised cluster, which we refer to as the set $\kappa$. Let $\Gamma=(V,E_\chi)$ be the undirected graph whose vertices correspond to pseudo-levels and where an edge exists between vertices $i$ and $j$ whenever $\chi_{ij} \geq \chi^*=1/2$. Then $\kappa$ is defined as the vertex set of the connected component of $\Gamma$ containing the first excited pseudo-level.

Let us now consider the case where the pseudo-ground state (from here on referred to as pGS) has negligible hybridisation with the rest of the pseudo-spectrum but we have a cluster of hybridised pseudo-eigenstates connected to the first pseudo-level.
Thus, in the present scenario we have $0 \notin \kappa$.
In this situation, we must first estimate bounds on the corrected pseudo-level due to the interaction with the rest of the cluster, and only then apply Eq.~\eqref{eq:bounds_gap}.

The interaction of the first pseudo-level with the rest of the cluster is accounted for by the effective interaction matrix $v_{\text{int}}$,
\begin{gather}
    (v_{\text{int}})_{ij} = \left\{ \begin{array}{cc}
        ||\nu_{ij}||_2 & i \neq j \\
        0 & i =j
    \end{array} \right.\  \forall i, j \in \kappa \ .
    \label{eq:v_int}
\end{gather}
The interaction with the cluster provides a new estimate for the pseudo-gap,
\begin{equation}
     \tilde{\Delta}_{10, l}^\prime = \tilde{\Delta}_{10} - ||v_{\text{int}}||_2 \ ,
\end{equation}
which becomes relevant to the lower bound.
On the other hand, the effective interaction between the pGS and a state of the cluster is upper-bounded by
\begin{equation}
    ||v_{\text{eff}}||_2 = \sqrt{\sum_{k \in \kappa} \left(||\nu_{0k}||_2\right)^2} \ .
\end{equation}
With these definitions, the bounds of the lowest spectral gap take the following shape in this case:
\begin{gather}
    \Delta_{10} \leq \tilde{\Delta}_{10} + \left\{
    \begin{array}{ll}
         \frac{2 \left(||\nu_{01}||_2\right)^2}{\tilde{\Delta}_{10} - ||\nu_{01}||_2} & \text{if } ||\nu_{01}||_2 < \tilde{\Delta}_{10} / 2 \\
        2 ||\nu_{01}||_2 & \text{otherwise}
    \end{array}
    \right. \ , 
    \label{eq:upper_bounds_gap_deg1exc} \\
    \Delta_{10} \geq \tilde{\Delta}^\prime_{10, l} - \left\{
    \begin{array}{ll}
         \frac{2 \left(||\nu_{\text{eff}}||_2\right)^2}{\tilde{\Delta}^\prime_{10, l} - ||\nu_{\text{eff}}||_2} & \text{if } ||\nu_{\text{eff}}||_2 < \tilde{\Delta}^\prime_{10, l} / 2 \\
        2 ||\nu_{\text{eff}}||_2 & \text{otherwise}
    \end{array}
    \right. \ . 
    \label{eq:lower_bounds_gap_deg1exc}
\end{gather}

The only scenario left to cover is $0\in \kappa$. In this case we cannot ensure that the gap doesn't close, and again the interactions within the cluster must be taken into account. Making use of Weyl's inequalities once more and of the min-max theorem, one can show that the bounds are given by
\begin{equation}
    0 \leq  \Delta_{10} \leq 
    ||v_{\text{int}}||_2 +
    \min_{j\in \kappa} \frac{\tilde{\Delta}_{j0}}{2} + \sqrt{\left( \frac{\tilde{\Delta}_{j0}}{2} \right)^2 + (||\nu_{0j}||_2)^2} 
    \label{eq:bounds_gap_clusterGS}
\end{equation}
in this high-hybridisation scenario. Notice that large hybridised clusters are expected to arise around a critical point, where the low-energy spectrum collapses towards a nearly degenerate manifold that becomes exactly degenerate in the thermodynamic limit.

\subsection{Feature inheritance of the pseudo-spectrum}\label{sec:qualitative_features}

In general, the bounds of~\cref{eq:bounds_gap,eq:lower_bounds_gap_deg1exc,eq:upper_bounds_gap_deg1exc,eq:bounds_gap_clusterGS} will be rather loose, especially in the presence of clusters, leaving us unable to rule out gap closings. However, can we still use the pseudo-eigenspectrum to learn something about their presence and, perhaps more interestingly, their nature?

Let us first consider the case where a first order phase transition takes place. In this scenario, since only two real eigenstates are involved in the true spectral gap closing, for reasonable CAS reductions (low $\varepsilon$) we should find a pseudo-gap closing in the vicinity of the true one, along with $\chi_{0k}, \chi_{1k} < 1/2 \ \forall k > 1$ around the area. Let us view this more in detail:
near the exponentially closing gap, the relevant subspace of $H(s)$ expressed in the pseudo-eigenstate basis becomes
\begin{equation}
    \hat{h}_{p, \text{eff}} = \begin{pmatrix}
        \mu_1 & \langle \mu_1 | \nu | \mu_0\rangle \\
        \langle \mu_0 | \nu | \mu_1\rangle & \mu_0
    \end{pmatrix} + \begin{pmatrix}
        \epsilon_1  & 0 \\
        0 & \epsilon_0
    \end{pmatrix} \ ,
    \label{eq:h_eff_1st_order}
\end{equation}
where $\epsilon_1 \ll \mu_1, \epsilon_0 \ll \mu_0$ represent the shift of the pseudo-eigenlevels that we have failed to account for by disregarding the interaction with the rest of the spectrum.
Due to the low-$\varepsilon$ assumption, we will find that the pseudo-gap can be described in the vicinity of the QPT as $\tilde{\Delta}_{10}(s) \simeq \alpha|\tilde{s}_{\min} - s|$ for some $\alpha > 0$, where $\tilde{s}_{\min}$ is the point where the pseudo-gap closes. Note that all the above conditions also imply that $|\langle \mu_1 | \nu | \mu_0\rangle| \ll 1$ and $|\epsilon_1 - \epsilon_0| \ll \alpha$.
Thus, the true gap becomes
\begin{equation}
    \Delta_{10} = \sqrt{[\alpha (\tilde{s}_{\min} - s) + (\epsilon_1 - \epsilon_0)]^2 + 4 |\langle \mu_1 | \nu | \mu_0\rangle|^2}  \ .
    \label{eq:relation_gap_pseudo-gap_1st_order}
\end{equation}
Considering the canonical expression of the hyperbola and assuming that the dominant $s$-dependence comes from the pseudo-level difference (which implies that the slopes $\alpha$ of the asymptotes of $\tilde{\Delta}$ are approximately equal to those of $\Delta$), we can identify $\Delta_{\min} \simeq 2|\langle \mu_1 | \nu | \mu_0\rangle|$ and $|s_{\min} - \tilde{s}_{\min}| \simeq \frac{|\epsilon_1 - \epsilon_0|}{\alpha}$.

In contrast, for the case of a critical region, present in continuous phase transitions, the true spectrum behaves rather differently; it is not only the first excited state but a tower of excited states that collapse onto the ground state. Thus, 
we expect to find a large, strongly hybridised cluster containing the pGS around the critical point,
such that $\chi_{0k}, \chi_{1k} > 1/2$ for many $k$. 

Another important characteristic of a continuous transition is the divergence of correlations near the critical point. This feature strongly suggests that the CAS mediating the first and ground pseudo-states (i.e., the one mediating the pseudo-gap closing) should be highly nonlocal in a low-error CAS reduction.

Thus, we can conclude that the pseudo-spectrum can serve as a powerful diagnostic for the presence of quantum phase transitions, as well as for their nature. Moreover, since the features of the phase transition will become sharper as we approach the thermodynamic limit, we expect the associated pseudo-spectral signatures to become increasingly well defined as system size grows. For sufficiently structured problems, the CAS framework is a tool that enables the fully analytical study of phase diagrams from a new perspective.


\section{A specific example: the adiabatic algorithm on an Ising model}\label{sec:TFIM_example}

To illustrate the capabilities of the CAS framework, we develop an in-depth treatment of the adiabatic interpolation between a transverse-field Hamiltonian and an Ising model and test it numerically. This interpolation can exhibit both first order transitions and a critical point, making it a useful scenario for the illustration of all the aspects of the CAS reduction described so far. This particular interpolation family is widely studied in the quantum annealing literature, where classical optimisation problems are mapped onto Hamiltonian interpolations.


For the remainder of this section we consider
\begin{gather}
    A = \frac{-1}{\sqrt{N}} \sum_i X_i \ ,
    \label{eq:A_TF_normalised}\\
    B = \frac{1}{\mathcal{N}_B} \left(\sum_i h^z_i Z_i + \sum_{j<i} J_{ij} Z_i Z_j \right) \ ,
    \label{eq:B_Ising_normalised}
\end{gather}
with $\mathcal{N}_B = \sqrt{\sum_i (h^z_i)^2 + \sum_{j<i} (J_{ij})^2}$. From here on, we absorb the normalisation constants into the values $h^x_i, h^z_j$ and $J_{ij}$.

Following the procedure introduced above, the problem reduces to identifying minimal term removals that induce a $\mathbb{Z}_2$ symmetry. Due to the restricted diversity of the alphabet $\Omega_{A\cup B} = \{X_i, Z_i, Z_iZ_j\}_{i, j< i}$, only four distinct CAS families arise. These admit compact analytic descriptions of the resulting projections and can be searched through efficiently, yielding a tractable reduction procedure. Their detailed characterisation and complexity analysis 
are deferred to Appendix~\ref{sec:CAS_reductions}; here we only present their essential features.

\begin{figure}
    \centering
    \includegraphics[width=\linewidth]{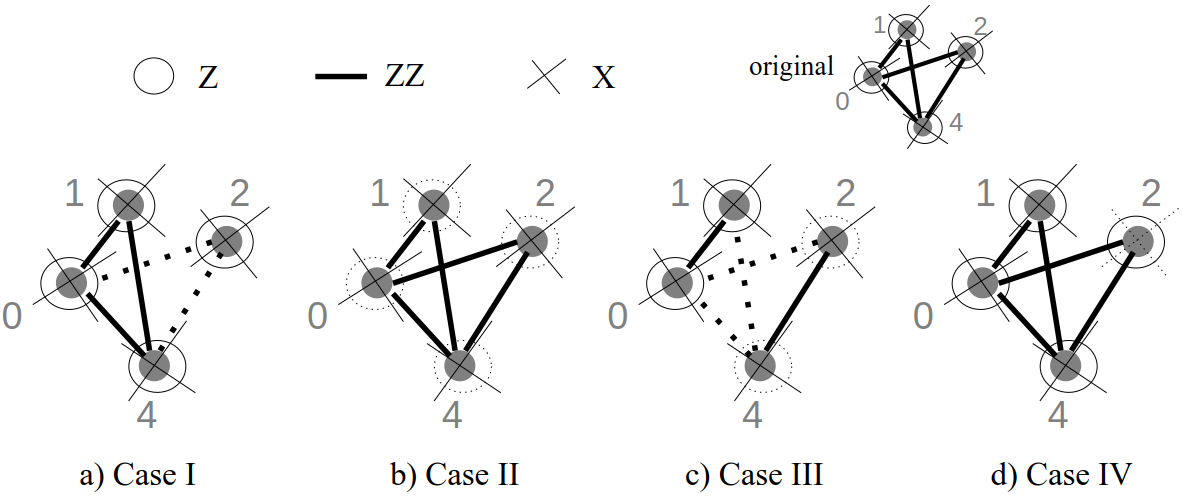}
    \caption{Illustration of the possible ways to generate a $\mathbb{Z}_2$ symmetry in $H(s)$ for $A, B$ of the shape in Eqs.~\eqref{eq:A_TF_normalised} and \eqref{eq:B_Ising_normalised}, respectively. Dotted lines represent the elimination of said terms (i.e., their inclusion in $\nu$), and the illustration labelled as ``original" is the one corresponding to the full representation of $H(s)$ as per Eq.~\eqref{eq:linear_interp_H}.}
    \label{fig:CAS_types}
\end{figure}

The resulting minimal types of term removals are illustrated in Fig.~\ref{fig:CAS_types} and can be summarised as follows:
\begin{enumerate}[I]
	\item Removal of all couplings $Z_iZ_k$ to a single qubit $k$. The induced symmetry $T_I\in \mathcal{T}_I$ is the (time-dependent) local field on said qubit. 

    There are $n_{\text{eff}} = N- r$ possible CAS of this type at recursion order $r$, each with a potentially different cost $\varepsilon$.
	\item Removal of all local fields in $z$. This produces the well-known parity symmetry of the transverse-field Ising model (TFIM), such that the induced symmetry $T_{II}\in \mathcal{T}_{II}$ is the product of all $X$ operators.

    There is a single possible CAS of this type, with a single cost $\varepsilon_{II}$, but a choice must be made in terms of the effective variable to trace over. Adopting a look-ahead strategy a single step into the future in order to choose it, searching over this CAS type remains polynomial in $n_{\text{eff}}$ at each recursion order, since the rest of CAS types also have a polynomial search cost in $n_{\text{eff}}$.
	\item Removal of all local fields in $z$ in a qubit subset and all couplings between said qubit subset and the rest of the system. This induces a parity symmetry in the aforementioned qubit subset, such that the induced symmetry $T_{III}\in \mathcal{T}_{III}$ is the product of all $X$ operators within the qubit subset. 

    The search for this CAS type is shown to remain polynomial in $n_{\text{eff}}$ after the appropriate relaxation, see Appendix~\ref{sec:caseIII}. Within the qubit subset, the reduction variable must be determined in a similar manner to case II, which is now a polynomial search in the size of the qubit subset $n'< n_{\text{eff}} -1 $. 
	\item Removal of the local $x$-field in a single qubit $k$. This makes the local $z$-field in the $k$-th qubit $T_{IV} = Z_k \in \mathcal{T}_{IV}$ a conserved quantity. 

    There are $n_{\text{eff}} = N- r$ possible CAS of this type at recursion order $r$, each a different cost $\varepsilon$ for inhomogeneous $x$-fields. In the homogeneous case of Eq.~\eqref{eq:A_TF_normalised}, there is a single value of the cost $\varepsilon_{IV}$ but the reduction variable must nonetheless be chosen, so the same single-step look-ahead strategy of case II and III can be employed if $\varepsilon_{IV}$ is minimal with respect to the rest of CAS types.
\end{enumerate}
The accessible CAS set is thus $\mathcal{T} = \{\mathcal{T}_I, \mathcal{T}_{II}, \mathcal{T}_{III}, \mathcal{T}_{IV}\}$, over which we can search for the minimum $\varepsilon$ cost in each recursion step performing a number of computations that is polynomial in $n_{\text{eff}}$.

We note that case I only corresponds to a full term removal at the beginning of the anneal (precisely when it is not present yet, as $s=0$), but since we can efficiently track the CAS basis throughout the interpolation we still consider it part of our accessible set $\mathcal{T}$. 

The CAS reduction pipeline is thus described by a recursion tree, where each node corresponds to a CAS search that decides the next Hilbert space splitting, and where the depth of each leaf corresponds to the recursion order. Every branch has a maximum depth of $N-1$, where the $2\times 2$ matrices are achieved, but there are exponentially many branches, so it is only tractable to follow $poly(N)$ of those. Once a CAS is chosen, we need to decide whether to stay on the $+1$ or on the $-1$ sector in order to find low-energy solutions. The explicit shape of each type of projection allows one to define a naïve criterion on which sector to keep for each CAS type (see Appendix~\ref{sec:CAS_reduction_complexity}), but hard problems, such as the ones found in nontrivial optimisations, will generally not follow it. Nonetheless, every set of rules to explore the CAS recursion tree defines a different heuristic to search for the low-energy space. In the specific case of interpolations between Eq.~\eqref{eq:A_TF_normalised} and Eq.~\eqref{eq:B_Ising_normalised}, each branch can be associated with a bitstring or a reduced set of bitstrings, such that the CAS reduction unlocks a new family of classical heuristics inspired by AQC.


\subsection{Numerical results}\label{sec:numerical_results}

\begin{figure}
    \centering
    \includegraphics[width=\linewidth]{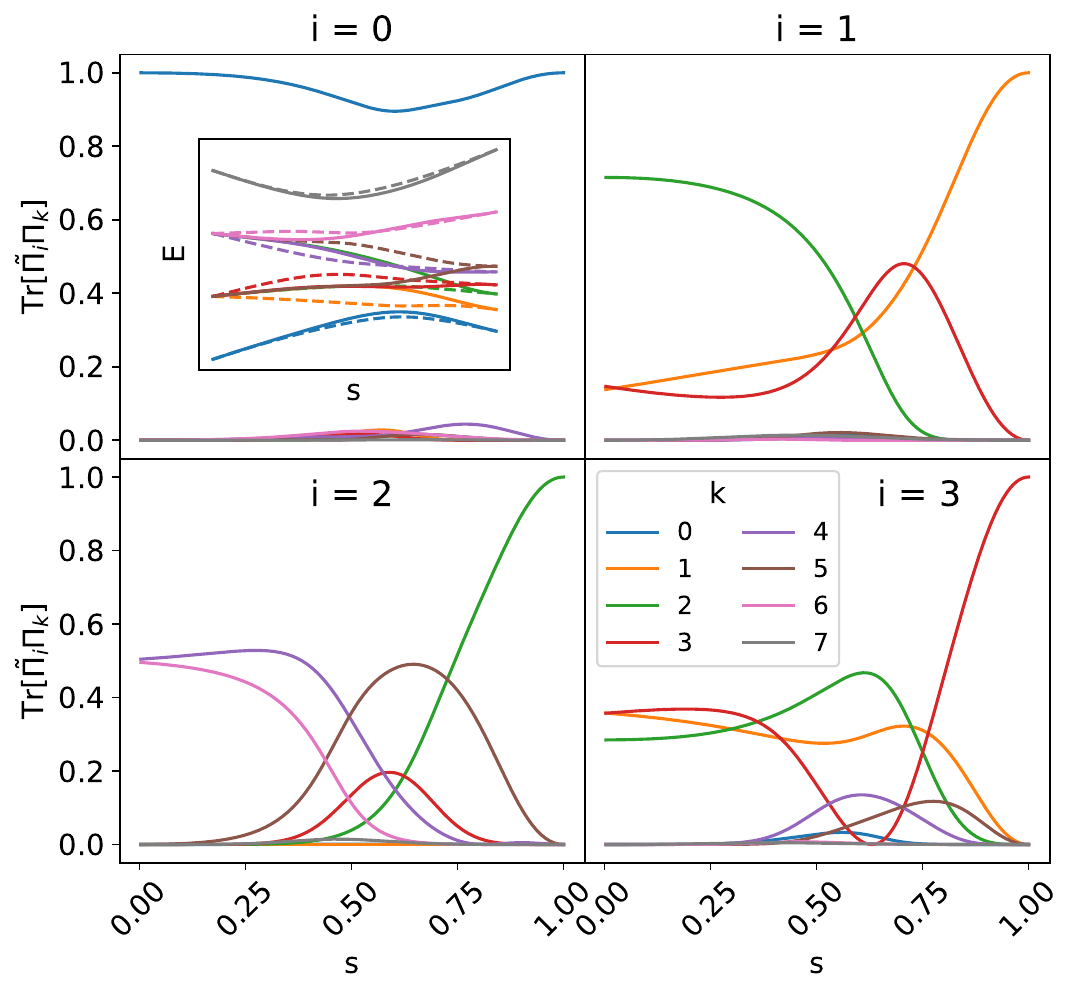}
  \caption{Fidelity between the first four pseudo-eigenspaces $\tilde{\Pi}_k$ and the true eigenspaces $\Pi_k$ of the system, where the pseudo-eigenspaces are labelled according to their ordering at $s=1$. The inset shows the true spectrum (dashed lines) and the pseudo-spectrum (solid) for this particular instance.}
  \label{fig:N3_pseudoeigenspaces}
\end{figure}

We first illustrate our approach on a small instance where we can compute the pseudo-eigenspaces, as well as all pseudo-levels, in order to provide a more complete picture of the approximation of the CAS reduction.
Fig.~\ref{fig:N3_pseudoeigenspaces} shows the instantaneous fidelities of the pseudo-eigenspaces with the true eigenspaces for a small $N=3$ random Ising problem. For this instance, where the pseudo-gap remains open, we see that the fidelity of the pGS with the GS remains quite large throughout the full interpolation. For the first, second and third pseudo-levels (ordered according to their energy at $s=1$), the situation is more complex. Pseudo-levels $\tilde{\Pi}_1, \tilde{\Pi}_3$ and $\tilde{\Pi}_5$ (the latter not shown) remain degenerate at low $s$, and thus we find that they are linear combinations within the true first excited eigenspace $\Pi_1 + \Pi_2 + \Pi_3$. A similar situation takes place for the second excited manifold at low $s$, from which we only show $\tilde{\Pi}_2$. However, when more significant changes in the true eigenstates take place the pseudo-eigenvalues start to cross, leading to shifts in the support over the true eigenspaces. Importantly, one can observe that the spread of this support remains fairly limited.

\begin{figure}
    \centering
    \includegraphics[width=0.50\linewidth]{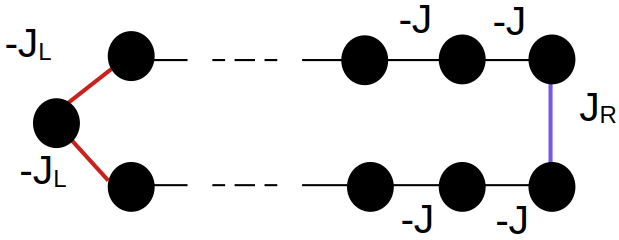}
    \caption{Schematic representation of the frustrated Ising ring model.}
    \label{fig:frustrated_ising_ring}
\end{figure}

Further small, pedagogical examples may be found in Appendix~\ref{sec:small_sizes}; we now proceed to test our approach in large system sizes.
For this we turn to the frustrated Ising ring model~\cite{roberts_noise_2020}, illustrated in Fig.~\ref{fig:frustrated_ising_ring}. This toy model consists on a closed chain of of $N=2k + 1$ spins with $k>5 \in \mathbb{N}_+$, all connected by equal couplings $J_{ij}=-J$ except for $J_{N-1,0} = J_R$ and $J_{(N+1)/2, (N-1)/2} = J_{(N-1)/2, (N-1)/2 - 1} = -J_L$, with $J_R < J_L < J$. The frustrated Ising ring is exactly solvable and it features well characterised continuous and first order phase transitions. In addition, the first order transition is a perturbative anticrossing happening towards the end of the interpolation between the final GS, where the only violated bond is $J_R$, and the first excited state, where one of the $J_L$ bonds is violated. Since the Hamming distance between both states is $(N-1)/2$, the transition matrix between them is indeed exponentially suppressed.

For this analysis we will consider $N=101$, so that we are reasonably close to the thermodynamic limit. We compute the $N+1=102$ CAS branches with the lowest energy, which due to the simple structure of the problem at hand are easy to identify and capture all the relevant low-energy physics. The technical criterion to follow a branch is to allow a single deviation from the naïve criterion that signals the low-energy block. In the language of Appendix~\ref{sec:CAS_reduction_complexity}, we restrict to paths with $K \leq 1$ choices that seem to strive away from lower energies. 
In order to run these large simulations we discretise time, since symbolic operations slow down the computation greatly due to the nested $s$-dependencies that arise from type I reductions, which can be found in Appendix~\ref{sec:caseI}. 

As ground truth to benchmark against, we have computed the first eight levels via exact diagonalisation~\cite{roberts_noise_2020} and find that the critical point where the continuous phase transition takes place is $s_c = 0.5135 \pm 0.0005$. The first order phase transition, on the other hand, takes place at $s_{\min} = 0.8979 \pm 0.0005$.

\begin{figure*}
\centering
    \begin{tikzpicture}
      \node[anchor=south west, inner sep=0] (img) at (0,0)
        {\includegraphics[width=\linewidth]{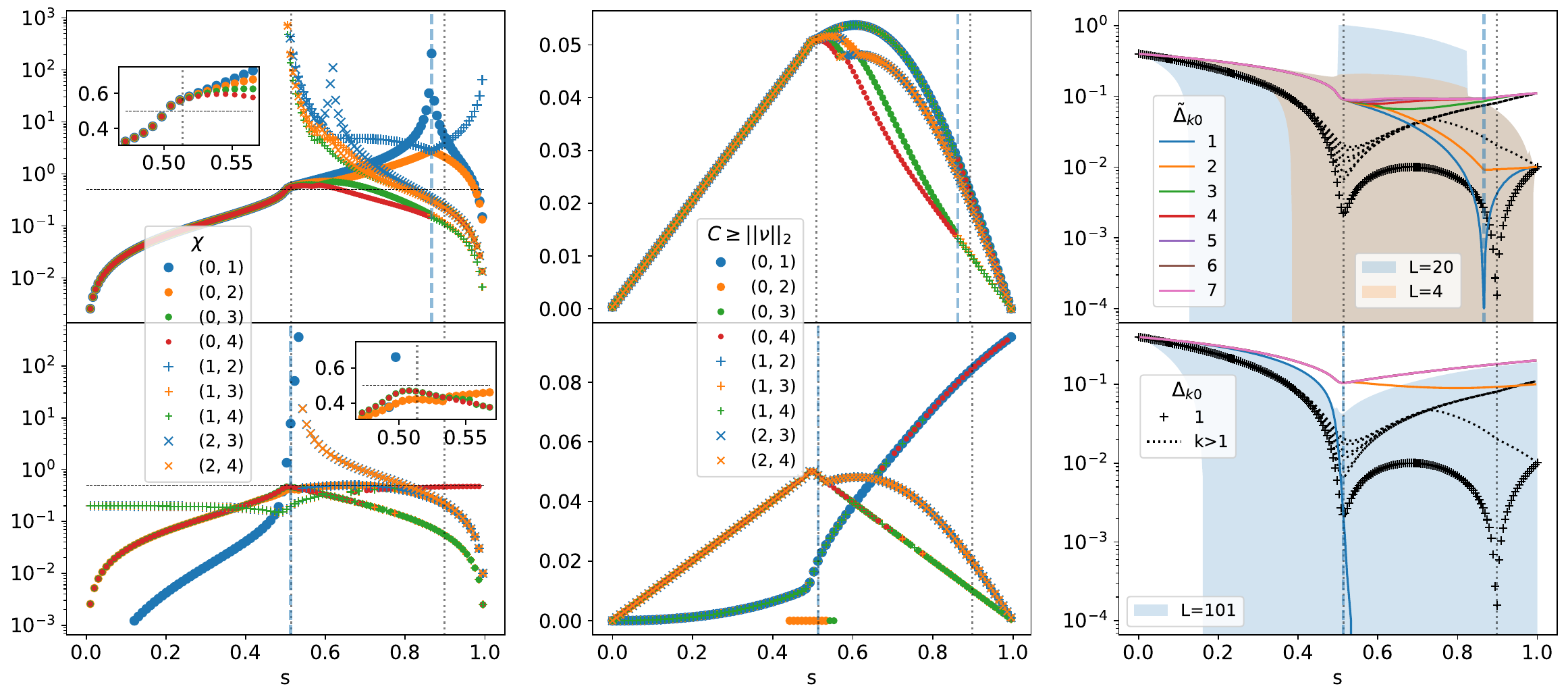} };
      \begin{scope}[x={(img.south east)}, y={(img.north west)}]
        \node[] at (0.07,0.94) {a)};
        \node[] at (0.41,0.94) {c)};
        \node[] at (0.75,0.95) {e)};
        \node[] at (0.08,0.16) {b)};
        \node[] at (0.41,0.19) {d)};
        \node[] at (0.75,0.19) {f)};
      \end{scope}
    \end{tikzpicture}
  \caption{Simulation of the $N=101$ frustrated Ising ring described in the main text without any biasing weights (top row, a, c, e) and with a bias towards minimising the error around the middle according to Eq.~\eqref{eq:nu_max_cost_weighted} (bottom row, b, d, f). The vertical black dotted lines mark $s_c$ and $s_{\min}$, and the vertical blue dashed lines indicate their respective predictions from the pseudo-spectra ($s_{\min}$ on the top row, $s_c$ on the bottom one). Plots a), b) show the hybridisations for all the pairs of the lowest four pseudo-levels for every $s$, where the horizontal dashed line indicates $\chi^*=0.5$ and we do not show values above $10^3$ or below $10^{-3}$ to improve visibility. The insets in a), b) present a zoom-in onto the hybridisations with the pGS, $\chi_{0k}$, near $s_c$. The off-diagonal norms for the lowest four pseudo-levels at each value of $s$ are shown in c) and d). Plots e) and f) present the first seven energy separations with the true GS (black) and with the pGS (colours) found for each cost function, where we set to zero values below $10^{-4}$ to aid visibility. The shaded regions indicate the gap bounds calculated considering the lowest $L$ pseudo-levels. We highlight that the occasional discontinuities in a)-d) are an artifact of pseudo-level degeneracy.
  }
  \label{fig:frustrated_ising_ring_N101_panel}
\end{figure*}
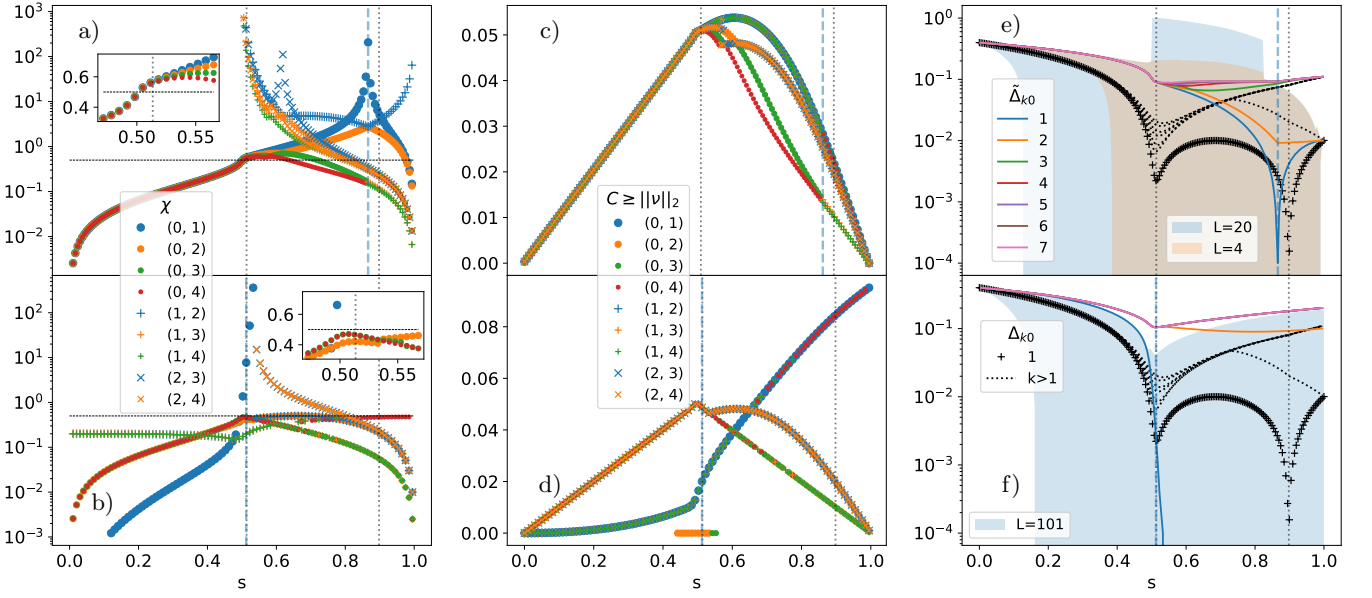

We find that the standard CAS reduction, performed minimising $\varepsilon$ from Eq.~\eqref{eq:nu_max_cost} provides a good description of the first order transition and shows signatures of the continuous one. However, since these signatures are not enough to provide a quantitative estimate of the critical point, we examine an alternative CAS reduction that is biased towards obtaining a lower error around the middle of the interpolation. The results of these simulations are shown in the top and bottom rows of Fig.~\ref{fig:frustrated_ising_ring_N101_panel}, respectively, which we now proceed to discuss in detail.

%
%
%

The results obtained when running the CAS reduction by minimising Eq.~\eqref{eq:nu_max_cost} are presented in the top row of Fig.~\ref{fig:frustrated_ising_ring_N101_panel}. We highlight that the indices of the pseudo-levels indicated in the legends refer to the absolute ordering of the said pseudo-levels at each value of $s$ throughout this figure. 
Excluding the trivial case II CAS at first recursion order, which is exact, the solution returned by the optimiser contains only type-I CAS, which are local and introduce zero error at the beginning and at the end of the interpolation. In addition, in this situation we have that the computed upper bound on the 2-norm of the off-diagonals exactly coincides with the true 2-norm ($C=||\nu||_2$). The obtained pseudo-gap provides a reasonably good prediction of the first order transition's location, $s_{\min}^{\text{pred}} = 0.8662 \pm 0.0015$, resulting in a relative error of 3.5\%. For visual comparison, $s_{\min}^{\text{pred}}$ is depicted as a blue dashed line in Fig.~\ref{fig:frustrated_ising_ring_N101_panel}a, ~\ref{fig:frustrated_ising_ring_N101_panel}c and ~\ref{fig:frustrated_ising_ring_N101_panel}e, whereas the black dotted line closest to it marks the true $s_{\min}$. When we look into the hybridisation around $s_{\min}$ in Fig.~\ref{fig:frustrated_ising_ring_N101_panel}a, we observe that we actually have a hybridised cluster comprising the ground, first and second pseudo-levels, which is likely to be the main error source for the displacement of $s_{\min}^{\text{pred}}$ with respect to $s_{\min}$. 

We highlight that the relevant off-diagonal for the crossing of $s_{\min}^{\text{pred}}$ arises at recursion order $r=(N+1)/2$, as numerically verified with the simulation of different system sizes. Such scaling of the order $r$ of the relevant CAS splitting with system size suggests a connection to the exponential suppression of the effective coupling between the relevant pseudo-eigenstates. Indeed, increasing $r$ corresponds to projecting onto progressively smaller sub-blocks, such that the associated $\nu^{(r)}$ connect increasingly reduced sectors of Hilbert space. It is worth noting that, due to the normalisation of $A$ and $B$, the total off-diagonal weight remains bounded throughout the reduction. Under the assumption that this weight is not strongly concentrated onto a small number of matrix elements in the pseudo-eigenbasis, the typical tunnelling matrix element mediating the interaction between pseudo-levels is expected to decrease with the size of the corresponding off-diagonal block. In the frustrated ring, the relevant mediating block has dimension $2^{N-r} = 2^{(N-1)/2}$. Interestingly, $(N-1)/2$ is precisely the Hamming distance separating the GS and first excited state. In this manner, the pseudo-spectrum analysis is consistent with the known exponential suppression of the gap at a perturbative anticrossing as a function of the Hamming distance between the states involved in such a first order transition.

In Fig.~\ref{fig:frustrated_ising_ring_N101_panel}e we also present the bounds obtained looking for clusters in up to the lowest $L$ pseudo-levels. For $s<s_c$, it was checked that $L=20$ provided almost indistinguishable results from the case where we consider the full first excited pseudo-manifold. Nonetheless, $L=4$ falls short as a good representative of the cluster and thus presents violations of the lower bound at low $s$. For $s_c - \epsilon \lesssim s \lesssim s_{\min}^{\text{pred}} - \epsilon^\prime$ for $\epsilon, \epsilon^\prime > 0$ we find a fully hybridised cluster containing all pseudo-levels, which makes the upper bound extensive on $L$. However, we see that in this regime $L=4$ suffices to provide a reliable upper bound. Both predictions coincide for $s \gtrsim s_{\min}^{\text{pred}} - \epsilon^\prime$ because the relevant cluster involves only the three lowest pseudo-levels in this region.

%


All in all, one may conclude that the standard CAS reduction provides a reasonably good description of the first order transition. However, the pseudo-gap presents no minimum around $s_c$, decreasing monotonically until $s_{\min}^{\text{pred}}$ instead. Nonetheless, it is interesting to note that the degeneracy of the pseudo-levels $\ket{\mu_{k>0}}$ begins to break shortly before $s_c$. This feature, if not appreciable with high precision in Fig.~\ref{fig:frustrated_ising_ring_N101_panel}e, is evidenced by the drop in hybridisation of the pairs of excited pseudo-levels in Fig.~\ref{fig:frustrated_ising_ring_N101_panel}a. This degeneracy, present for $s \lesssim s_c$, is indeed consistent with the paramagnetic phase of the transverse-field Ising model, where there is no magnetisation along the $z$-axis and thus the levels with the same total spin along the $x$-axis remain indistinguishable for pseudo-eigenstates made out of all-case-I reductions. 
In addition, the $\chi_{0k}$ go above the strongly hybridised threshold shortly before $s_c$, as we highlight in the inset of Fig.~\ref{fig:frustrated_ising_ring_N101_panel}a. As mentioned in the Results section, this behaviour is consistent with the presence of a critical point in the vicinity, since several excited levels approach the ground state in the true spectrum (see Figs.~\ref{fig:frustrated_ising_ring_N101_panel}e and~\ref{fig:frustrated_ising_ring_N101_panel}f).

However, none of these features provide a concrete estimate of the location of the critical point; they're mere indicators of its presence. Indeed, in Fig.~\ref{fig:frustrated_ising_ring_N101_panel}c one can note that the true $s_c$ is located in the region where the error $\varepsilon \sim ||\nu||_2$ is maximum, which is preventing us from obtaining more precise information on this transition. In order to surpass this limitation, we consider the weighted cost $\varepsilon_w$, 
\begin{equation}
    \varepsilon_w = \max_s\  w(s) \cdot ||\nu(s)||_2 \ ,
    \label{eq:nu_max_cost_weighted}
\end{equation}
with weights
\begin{equation}
    w(s) = \left\{
    \begin{array}{cc}
        3s & 0\leq s\leq 1/3 \\
        1 & 1/3 < s < 2/3 \\
        3(1-s) & 2/3 \leq s \leq 1
    \end{array} \right. \ ,
    \label{eq:weights4critical}
\end{equation}
in which we bias towards better accuracies around the middle of the interpolation.
The results of performing the CAS reduction minimising~\eqref{eq:nu_max_cost_weighted} are presented on the bottom row of Fig.~\ref{fig:frustrated_ising_ring_N101_panel}. In contrast to the unbiased case, we now find a case II in all $K=1$ branches at order $r=(N+1)/2$, where we have again checked the scaling with different system sizes. The rest of CAS reductions, apart from the trivial case II due to the exact $\mathbb{Z}_2$ symmetry, are of type I once again. 

The labels of the two lowest pseudo-levels throughout the interpolation are $\tilde{\xi}^0$ and $\tilde{\xi}^{1}_{51}$, as in the standard CAS reduction, but this time the first excited pseudo-level, $\tilde{\xi}^{1}_{51}$ breaks away from the first excited manifold already at low $s$. This can most clearly be observed in Fig.~\ref{fig:frustrated_ising_ring_N101_panel}b, where the hybridisations $\chi_{12}, \chi_{13}$ and $\chi_{14}$ are shown to be weak. As shown in Fig.~\ref{fig:frustrated_ising_ring_N101_panel}f, the first excited pseudo-level approaches the pGS exponentially fast shortly before the critical point, generating a degenerate pseudo-ground manifold afterwards. Note in Fig.~\ref{fig:frustrated_ising_ring_N101_panel}d that $||\nu_{01}||_2$ remains relatively small until around $s_c$ but grows notably past it, pointing to the fact that said degeneracy is an artifact of disregarding the interaction between both pseudo-levels. It is worth emphasising that the mediating CAS between these two pseudo-levels is of type II, that is, highly nonlocal, which as mentioned in the Results section is what we would expect for a continuous phase transition if the discarded $\nu$ is small enough. 

In contrast to the unbiased case, it was really necessary to consider the full first excited pseudo-manifold in order to obtain reliable lower bounds, as shown in Fig.~\ref{fig:frustrated_ising_ring_N101_panel}f. However, since for this CAS reduction there never is a large hybridised cluster that contains the ground state (see the inset in Fig.~\ref{fig:frustrated_ising_ring_N101_panel}b), the upper bound does not grow with $L$. In turn, the uncertainty at high $s$ is now much greater, since as shown in Fig.~\ref{fig:frustrated_ising_ring_N101_panel}e the error is maximal at the end of the interpolation.

Since the pseudo-gap simply closes in this CAS reduction, we turn our attention to the gap between the pGS and the higher excited pseudo-levels to determine a concrete prediction of the critical point. As the overall error is now lower around this region and, consequently, the representation of the true eigenvalues given by the pseudo-levels is more faithful, this time we do find that the $\tilde{\Delta}_{k>1, 0}$ curves present a dip around $s_c$. In fact, for the lowest 60 pseudo-levels found (above the first excited one), the minimum is found at the same point within simulation precision, $s_c^{\text{pred}} = 0.5125 \pm 0.0025$, which is exact within the present discretisation. 

Note that throughout the bottom row of Fig.~\ref{fig:frustrated_ising_ring_N101_panel}, the pseudo-spectrum obtained from minimising $\varepsilon_w$ contains no information about the first order transition; we traded it off for a more accurate description of the critical region.

\section{Conclusions and outlook}\label{sec:conclusions}

The CAS framework we have introduced here provides a novel approach to the analysis of interpolation paths in Hamiltonian space. By recursively slicing the Hilbert space in half in an imperfect but certifiably proper manner and such that the discarded off-diagonal is minimised, we can decompose the problem into a compressed but imperfect description of the interpolation process plus a structured error term.
We may identify the pseudo-spectrum with a compressed representation of the original interpolation problem because we are able to describe each branch, corresponding to a pseudo-level, by a string of $N$ CAS operators, which by working assumption must be efficiently representable, along with the relative sign of the sectors onto which we projected along the way. This decomposition allows us to identify properties of the true interpolation, such as the presence or absence of a phase transition, its nature, and can even provide some quantitative predictions as to their location. 

As a concrete example of the potential of the CAS decomposition, we have explicitly derived the CAS family for interpolations involving arbitrary Ising models on a transverse field. By applying this theory, we efficiently simulate the low-energy spectrum of the frustrated Ising ring near the thermodynamic limit ($N=101$). In this benchmark problem, we have been able to obtain reasonably accurate predictions of the locations of the two phase transitions present, of $3.5\%$ for the first order transition and below $0.5\%$ for the continuous transition. In addition, we observe that the recursion order of the split that mediates the pseudo-level crossing that represents the first order transition produces an estimate of the minimum gap size that is consistent with the true value.




The usefulness of the CAS reduction
stems from the fact that the recursive minimisation of the off-diagonal blocks fosters the localisation of the pseudo-eigenbasis in the true eigenbasis. Indeed, small off-diagonal couplings suppress the hybridisation between different pseudo-spaces, such that each pseudo-eigenstate is expected to retain support only over a restricted subset of true eigenstates.

In view of the aforementioned localisation properties, the compelling possibility of having a polynomial support of the pseudo-eigenspaces in the true eigenbasis arises. Establishing conditions under which the support of pseudo-eigenstates remains bounded or grows subexponentially is beyond the scope of this work, but we highlight it as an interesting open problem. Nonetheless, the physical picture of the CAS reduction presented so far motivates the question of whether suitable CAS families can yield pseudo-eigenspaces with support over the true eigenbasis that remains polynomial in system size, at least away from criticality.

Within this picture, another compelling open question arises: if a pseudo-state associated to some CAS recursion branch has a relatively sparse support in the true eigenbasis, one can consider whether a tensor network that is based on the effective degrees of freedom singled out by the CAS projections will require a lower bond dimension to faithfully represent the true eigenstate it is closest to. This integration of the CAS framework with tensor networks would then imply an additional layer involving the ``thermalisation" of the time-dependent effective degrees of freedom themselves at each time step. The conditions that guarantee a spread that remains polynomial in system size within this framework could potentially have novel implications in terms of classical simulability.

The CAS reduction provides a fully analytical description of the approximation found to the true spectrum, along with its associated uncertainties $\nu$. This may be readily exploited in some highly structured models, but in practice the discretisation of time is required to tackle large system sizes, as mentioned in the numerical results section. This is a consequence of the increasingly nested expressions that arise as recursion order increases, which critically slow down symbolic calculus operations. However, we highlight that the cost of reducing the time step, i.e., refining the resolution of the grid, is rather low in our setup, as we are simply increasing the dimension of the vectors to operate on linearly in grid size. In addition, the operations that we need to perform on these are elementary; subtractions, additions and max-searches. In this manner, not only does the CAS framework remove the need for time discretisation conceptually but also greatly reduces its computational burden in practice, enabling tractable, high-precision computation of the pseudo-spectrum.

Regarding tractability, we demonstrate that that the calculation of a polynomial number of pseudo-levels
can be done in polynomial time for a sufficiently manageable CAS set. From here, the main challenge that remains is identifying the branches that participate in the low-energy subspaces; namely ensuring that the branch connecting to the GS on the more complex Hamiltonian ($B$, in the standard AQC setting) is included within the polynomial number of branches we explore. 
Indeed, within the CAS framework, the hardness of tracking the low-energy subspace is expressed through the pGS branch at $s=1$ as the number of times we have deviated from the naïve choice of low-energy sector. 
This measures the degree of surprisal that said pseudo-state ended up having the lowest energy. We argue that this provides a natural notion of hardness of an interpolation problem; one which is built on the framework of adiabatic computation but remains agnostic to the adiabatic algorithm itself.


Finally, we expect that the CAS reduction can readily serve as a useful tool for the design and study of adiabatic algorithms, as it provides a general framework not only to examine different problem classes, but also for the study of catalyst terms, for example. In the specific context of classical optimisation, the decomposition with respect to the studied accessible set $\mathcal{T}$ unlocks a whole new family of classical heuristics,
where the common factor is a fundamental connection to the physics of Hamiltonian interpolation.




\section*{Acknowledgments}
We thank Matthias Werner for introducing us to the frustrated Ising ring model and kindly providing us with the code to compute its exact energy spectrum. In addition, we thank David Pérez-García for insightful discussions.

\section*{Code availability}
The custom code developed for this project can be found at \url{https://github.com/anapalu/closest-accessible-symmetries}.

\bibliography{references} 

@article{zanardi_information-theoretic_2007,
	title = {Information-{Theoretic} {Differential} {Geometry} of {Quantum} {Phase} {Transitions}},
	volume = {99},
	url = {https://link.aps.org/doi/10.1103/PhysRevLett.99.100603},
	doi = {10.1103/PhysRevLett.99.100603},
	number = {10},
	urldate = {},
	journal = {Physical Review Letters},
	author = {Zanardi, Paolo and Giorda, Paolo and Cozzini, Marco},
	month = sep,
	year = {2007},
	note = {Publisher: American Physical Society},
	pages = {100603},
}

@article{rezakhani_intrinsic_2010,
	title = {Intrinsic geometry of quantum adiabatic evolution and quantum phase transitions},
	volume = {82},
	url = {https://link.aps.org/doi/10.1103/PhysRevA.82.012321},
	doi = {10.1103/PhysRevA.82.012321},
	number = {1},
	urldate = {},
	journal = {Physical Review A},
	author = {Rezakhani, A. T. and Abasto, D. F. and Lidar, D. A. and Zanardi, P.},
	month = jul,
	year = {2010},
	note = {Publisher: American Physical Society},
	pages = {012321},
}

@article{albash_adiabatic_2018,
	title = {Adiabatic quantum computation},
	volume = {90},
	issn = {0034-6861, 1539-0756},
	url = {https://link.aps.org/doi/10.1103/RevModPhys.90.015002},
	doi = {10.1103/RevModPhys.90.015002},
	number = {1},
	urldate = {},
	journal = {Reviews of Modern Physics},
	author = {Albash, Tameem and Lidar, Daniel A.},
	month = jan,
	year = {2018},
	pages = {015002},
}

@article{born_beweis_1928,
  title     = "Beweis des Adiabatensatzes",
  author    = "Born, M and Fock, V",
  journal   = "Eur. Phys. J. A",
  publisher = "Springer Nature",
  volume    =  51,
  number    = "3-4",
  pages     = "165--180",
  month     =  mar,
  year      =  1928,
}

@article{kato_adiabatic_1950,
  title     = "On the adiabatic theorem of quantum mechanics",
  author    = "Kato, Tosio",
  journal   = "J. Phys. Soc. Jpn.",
  publisher = "Physical Society of Japan",
  volume    =  5,
  number    =  6,
  pages     = "435--439",
  month     =  nov,
  year      =  1950,
}

@article{amin_consistency_2009,
	title = {Consistency of the {Adiabatic} {Theorem}},
	volume = {102},
	url = {https://link.aps.org/doi/10.1103/PhysRevLett.102.220401},
	doi = {10.1103/PhysRevLett.102.220401},
	number = {22},
	urldate = {},
	journal = {Physical Review Letters},
	author = {Amin, M. H. S.},
	month = jun,
	year = {2009},
	note = {Publisher: American Physical Society},
	pages = {220401},
}

@article{jansen_bounds_2007,
  title     = "Bounds for the adiabatic approximation with applications to
               quantum computation",
  author    = "Jansen, Sabine and Ruskai, Mary-Beth and Seiler, Ruedi",
  journal   = "J. Math. Phys.",
  publisher = "AIP Publishing",
  volume    =  48,
  number    =  10,
  pages     = "102111",
  month     =  oct,
  year      =  2007,
}

@article{comparat_general_2009,
  title = {General conditions for quantum adiabatic evolution},
  author = {Comparat, Daniel},
  journal = {Phys. Rev. A},
  volume = {80},
  issue = {1},
  pages = {012106},
  numpages = {7},
  year = {2009},
  month = jul,
  publisher = {American Physical Society},
  doi = {10.1103/PhysRevA.80.012106},
  url = {https://link.aps.org/doi/10.1103/PhysRevA.80.012106}
}

@article{bachmann_adiabatic_2017,
  title = {Adiabatic Theorem for Quantum Spin Systems},
  author = {Bachmann, S. and De Roeck, W. and Fraas, M.},
  journal = {Phys. Rev. Lett.},
  volume = {119},
  issue = {6},
  pages = {060201},
  numpages = {6},
  year = {2017},
  month = aug,
  publisher = {American Physical Society},
  doi = {10.1103/PhysRevLett.119.060201},
  url = {https://link.aps.org/doi/10.1103/PhysRevLett.119.060201}
}

@article{wiersema_classification_2024,
	title = {Classification of dynamical {Lie} algebras of 2-local spin systems on linear, circular and fully connected topologies},
	volume = {10},
	copyright = {2024 The Author(s)},
	issn = {2056-6387},
	url = {https://www.nature.com/articles/s41534-024-00900-2},
	doi = {10.1038/s41534-024-00900-2},
	number = {1},
	urldate = {},
	journal = {npj Quantum Information},
	author = {Wiersema, Roeland and Kökcü, Efekan and Kemper, Alexander F. and Bakalov, Bojko N.},
	month = nov,
	year = {2024},
	note = {Publisher: Nature Publishing Group},
	pages = {110},
}

@article{kokcu_classification_2024,
	title = {Classification of dynamical {Lie} algebras generated by spin interactions on undirected graphs},
	url = {https://arxiv.org/abs/2409.19797},
	journal = {arXiv.org},
	author = {Kokcu, Efekan and Wiersema, Roeland and Kemper, Alexander F and Bakalov, Bojko N},
	month = aug,
	year = {2024},
	note = {arXiv:2409.19797v1},
}

@article{aguilar_full_2024,
	title = {Full classification of Pauli Lie algebras},
	url = {https://arxiv.org/abs/2408.00081},
	journal = {arXiv.org},
	author = {Aguilar, Gerard and Cichy, Simon and Eisert, Jens and Bittel, Lennart},
	month = jul,
	year = {2024},
	note = {arxiv:2408.00081},
}

@article{fujii_eigenvalue-invariant_2023,
	title = {Eigenvalue-{Invariant} {Transformation} of {Ising} {Problem} for {Anti}-{Crossing} {Mitigation} in {Quantum} {Annealing}},
	volume = {92},
	issn = {0031-9015},
	url = {https://journals.jps.jp/doi/10.7566/JPSJ.92.044001},
	doi = {10.7566/JPSJ.92.044001},
	number = {4},
	urldate = {},
	journal = {Journal of the Physical Society of Japan},
	author = {Fujii, Toru and Komuro, Koshi and Okudaira, Yosuke and Sawada, Masayasu},
	month = apr,
	year = {2023},
	note = {Publisher: The Physical Society of Japan},
	pages = {044001},
}

@article{palacios_scalable_2025,
	title = {Scalable 2-local architecture for quantum annealing of {Ising} models with arbitrary dimensions},
	volume = {23},
	url = {https://link.aps.org/doi/10.1103/PhysRevApplied.23.054070},
	doi = {10.1103/PhysRevApplied.23.054070},
	number = {5},
	urldate = {},
	journal = {Physical Review Applied},
	author = {Palacios, Ana and Garcia-Saez, Artur and Juliá-Díaz, Bruno and Estarellas, Marta P.},
	month = may,
	year = {2025},
	note = {Publisher: American Physical Society},
	pages = {054070},
}

@book{bhatia_matrix_1997,
	address = {New York, NY},
	series = {Graduate {Texts} in {Mathematics}},
	title = {Matrix {Analysis}},
	volume = {169},
	copyright = {http://www.springer.com/tdm},
	isbn = {978-1-4612-0653-8},
	url = {http://link.springer.com/10.1007/978-1-4612-0653-8},
	urldate = {},
	publisher = {Springer},
	author = {Bhatia, Rajendra},
	year = {1997},
	doi = {10.1007/978-1-4612-0653-8},
}

@book{kato_perturbation_1995,
	address = {Berlin, Heidelberg},
	series = {Classics in {Mathematics}},
	title = {Perturbation {Theory} for {Linear} {Operators}},
	volume = {132},
	copyright = {http://www.springer.com/tdm},
	isbn = {978-3-642-66282-9},
	url = {http://link.springer.com/10.1007/978-3-642-66282-9},
	urldate = {},
	publisher = {Springer},
	author = {Kato, Tosio},
	year = {1995},
	doi = {10.1007/978-3-642-66282-9},
}

@book{ahuja_network_1993,
	address = {Upper Saddle River, NJ},
	title = {Network flows : theory, algorithms, and applications},
	isbn = {978-0-13-617549-0},
	shorttitle = {Network {Flows}},
	url = {http://archive.org/details/networkflowstheo0000ahuj},
	publisher = {Pearson},
	author = {Ahuja, Ravindra and Magnanti, Thomas and Orlin, James},
	year = {1993},
}

@inproceedings{orlin_max_2013,
author = {Orlin, James B.},
title = {Max flows in O(nm) time, or better},
year = {2013},
isbn = {9781450320290},
publisher = {Association for Computing Machinery},
address = {New York, NY, USA},
url = {https://doi.org/10.1145/2488608.2488705},
doi = {10.1145/2488608.2488705},
booktitle = {Proceedings of the Forty-Fifth Annual ACM Symposium on Theory of Computing},
pages = {765–774},
numpages = {10},
location = {Palo Alto, California, USA},
series = {STOC '13}
}

@article{roberts_noise_2020,
	title = {Noise amplification at spin-glass bottlenecks of quantum annealing: {A} solvable model},
	volume = {101},
	shorttitle = {Noise amplification at spin-glass bottlenecks of quantum annealing},
	url = {https://link.aps.org/doi/10.1103/PhysRevA.101.042317},
	doi = {10.1103/PhysRevA.101.042317},
	number = {4},
	urldate = {},
	journal = {Physical Review A},
	author = {Roberts, David and Cincio, Lukasz and Saxena, Avadh and Petukhov, Andre and Knysh, Sergey},
	month = apr,
	year = {2020},
	note = {Publisher: American Physical Society},
	pages = {042317},
}

@article{zanardi_ground_2006,
	title = {Ground state overlap and quantum phase transitions},
	volume = {74},
	url = {https://link.aps.org/doi/10.1103/PhysRevE.74.031123},
	doi = {10.1103/PhysRevE.74.031123},
	number = {3},
	urldate = {},
	journal = {Physical Review E},
	author = {Zanardi, Paolo and Paunković, Nikola},
	month = sep,
	year = {2006},
	note = {Publisher: American Physical Society},
	pages = {031123},
}

@article{kumar_geodesics_2012,
	title = {Geodesics in information geometry: {Classical} and quantum phase transitions},
	volume = {86},
	issn = {1539-3755, 1550-2376},
	shorttitle = {Geodesics in information geometry},
	url = {https://link.aps.org/doi/10.1103/PhysRevE.86.051117},
	doi = {10.1103/PhysRevE.86.051117},
	language = {},
	number = {5},
	urldate = {},
	journal = {Physical Review E},
	author = {Kumar, Prashant and Mahapatra, Subhash and Phukon, Prabwal and Sarkar, Tapobrata},
	month = nov,
	year = {2012},
	pages = {051117},
}

@article{larocca_diagnosing_2022,
  doi = {10.22331/q-2022-09-29-824},
  url = {https://doi.org/10.22331/q-2022-09-29-824},
  title = {Diagnosing {B}arren {P}lateaus with {T}ools from {Q}uantum {O}ptimal {C}ontrol},
  author = {Larocca, Martin and Czarnik, Piotr and Sharma, Kunal and Muraleedharan, Gopikrishnan and Coles, Patrick J. and Cerezo, M.},
  journal = {{Quantum}},
  issn = {2521-327X},
  publisher = {{Verein zur F{\"{o}}rderung des Open Access Publizierens in den Quantenwissenschaften}},
  volume = {6},
  pages = {824},
  month = sep,
  year = {2022}
}

@article{fontana_characterizing_2024,
	title = {Characterizing barren plateaus in quantum ansätze with the adjoint representation},
	volume = {15},
	url = {https://www.nature.com/articles/s41467-024-49910-w},
	doi = {10.1038/s41467-024-49910-w},
	number = {1},
	urldate = {},
	journal = {Nature Communications},
	author = {Fontana, Enrico and Herman, Dylan and Chakrabarti, Shouvanik and Kumar, Niraj and Yalovetzky, Romina and Heredge, Jamie and Sureshbabu, Shree Hari and Pistoia, Marco},
	month = aug,
	year = {2024},
	note = {},
	pages = {7171},
}

@article{ragone_a_2024,
	title = {A Lie algebraic theory of barren plateaus for deep parameterized quantum circuits},
	volume = {15},
	url = {https://www.nature.com/articles/s41467-024-49909-3},
	doi = {10.1038/s41467-024-49909-3},
	number = {1},
	urldate = {},
	journal = {Nature Communications},
	author = {Ragone, Michael and Bakalov, Bojko N. and Sauvage, Frédéric and Kemper, Alexander F. and Ortiz Marrero, Carlos and Larocca, Martín and Cerezo, M.},
	month = aug,
	year = {2024},
	note = {},
	pages = {7172},
}

@book{dalessandro_introduction_2021,
  doi = {10.1201/9781003051268},
  title = {Introduction to Quantum Control and Dynamics},
  author = {D'Alessandro, Domenico},
  edition = {2},
  address = {Boca Raton},
  isbn = {978-1-00-305126-8},
  publisher = {Chapman and Hall/CRC},
  month = sep,
  year = {2021}
}

@article{goh_lie_2025,
  title = {Lie-algebraic classical simulations for quantum computing},
  author = {Goh, Matthew L. and Larocca, Martin and Cincio, Lukasz and Cerezo, M. and Sauvage, Fr\'ed\'eric},
  journal = {Phys. Rev. Res.},
  volume = {7},
  issue = {3},
  pages = {033266},
  numpages = {32},
  year = {2025},
  month = sep,
  publisher = {American Physical Society},
  doi = {10.1103/3y65-f5w6},
  url = {https://link.aps.org/doi/10.1103/3y65-f5w6}
}

@article{barligea_enabling_2026,
	title = {Enabling {Lie}-{Algebraic} {Classical} {Simulation} beyond {Free} {Fermions}},
	url = {http://arxiv.org/abs/2604.16701},
	doi = {10.48550/arXiv.2604.16701},
	publisher = {arXiv},
	author = {Bärligea, Adelina and Sims-Goh, Matthew L. and Kottmann, Jakob S.},
	month = may,
    journal = {arXiv.org},
	year = {2026},
	note = {arXiv:2604.16701 [quant-ph]},
}

@book{zhang_schur_2005,
  title     = {The schur complement and its applications},
  editor    = {Zhang, Fuzhen},
  publisher = {Springer},
  series    = {Numerical Methods and Algorithms},
  edition   =  {2005},
  month     =  mar,
  year      =  {2005},
  address   = {New York, NY},
  doi = {https://doi.org/10.1007/b105056},
  language  = {}
}

@article{griesemer_smooth_2008,
  title     = {On the smooth Feshbach--Schur map},
  author    = {Griesemer, M and Hasler, D},
  journal   = {J. Funct. Anal.},
  publisher = {Elsevier BV},
  volume    =  {254},
  number    =  {9},
  pages     = {2329--2335},
  month     =  may,
  year      =  {2008},
  doi = {https://doi-org.sire.ub.edu/10.1016/j.jfa.2008.01.015},
  language  = {}
}

@book{cardy_scaling_1996,
  title     = {Scaling and renormalization in statistical physics},
  author    = {Cardy, John L},
  publisher = {Cambridge University Press},
  month     =  apr,
  year      =  {1996},
  address   = {Cambridge, England},
  collection={Cambridge Lecture Notes in Physics},
  isbn = {9781316036440},
  doi = {10.1017/CBO9781316036440}
}

@article{fisher_disorderedRG_1995,
  title = {Critical behavior of random transverse-field Ising spin chains},
  author = {Fisher, Daniel S.},
  journal = {Phys. Rev. B},
  volume = {51},
  issue = {10},
  pages = {6411--6461},
  numpages = {0},
  year = {1995},
  month = mar,
  publisher = {American Physical Society},
  doi = {10.1103/PhysRevB.51.6411},
  url = {https://link.aps.org/doi/10.1103/PhysRevB.51.6411}
}

@InProceedings{pmlr-v32-kondor14,
  title = 	 {Multiresolution Matrix Factorization},
  author = 	 {Kondor, Risi and Teneva, Nedelina and Garg, Vikas},
  booktitle = 	 {Proceedings of the 31st International Conference on Machine Learning},
  pages = 	 {1620--1628},
  year = 	 {2014},
  editor = 	 {Xing, Eric P. and Jebara, Tony},
  volume = 	 {32},
  series = 	 {Proceedings of Machine Learning Research},
  address = 	 {Bejing, China},
  month = 	 jun,
  publisher =    {PMLR},
  url = 	 {https://proceedings.mlr.press/v32/kondor14.html}
}

@article{white_dmrg_1993,
  title = {Density-matrix algorithms for quantum renormalization groups},
  author = {White, Steven R.},
  journal = {Phys. Rev. B},
  volume = {48},
  issue = {14},
  pages = {10345--10356},
  numpages = {0},
  year = {1993},
  month = oct,
  publisher = {American Physical Society},
  doi = {10.1103/PhysRevB.48.10345},
  url = {https://link.aps.org/doi/10.1103/PhysRevB.48.10345}
}

@article{verstraete_dmrg_2004,
  title = {Density Matrix Renormalization Group and Periodic Boundary Conditions: A Quantum Information Perspective},
  author = {Verstraete, F. and Porras, D. and Cirac, J. I.},
  journal = {Phys. Rev. Lett.},
  volume = {93},
  issue = {22},
  pages = {227205},
  numpages = {4},
  year = {2004},
  month = nov,
  publisher = {American Physical Society},
  doi = {10.1103/PhysRevLett.93.227205},
  url = {https://link.aps.org/doi/10.1103/PhysRevLett.93.227205}
}

@article{paeckel_tebd_2019,
  title     = {Time-evolution methods for matrix-product states},
  author    = {Paeckel, Sebastian and K{\"o}hler, Thomas and Swoboda, Andreas
               and Manmana, Salvatore R and Schollw{\"o}ck, Ulrich and Hubig,
               Claudius},
  journal   = {Ann. Phys. (N. Y.)},
  publisher = {Elsevier BV},
  volume    =  {411},
  pages     = {167998},
  month     =  dec,
  year      =  {2019},
}

@article{cirac_mpsreview_2021,
  title = {Matrix product states and projected entangled pair states: Concepts, symmetries, theorems},
  author = {Cirac, J. Ignacio and P\'erez-Garc\'{\i}a, David and Schuch, Norbert and Verstraete, Frank},
  journal = {Rev. Mod. Phys.},
  volume = {93},
  issue = {4},
  pages = {045003},
  numpages = {65},
  year = {2021},
  month = dec,
  publisher = {American Physical Society},
  doi = {10.1103/RevModPhys.93.045003},
  url = {https://link.aps.org/doi/10.1103/RevModPhys.93.045003}
}

@article{wei_efficient_2023,
  title = {Efficient adiabatic preparation of tensor network states},
  author = {Wei, Zhi-Yuan and Malz, Daniel and Cirac, J. Ignacio},
  journal = {Phys. Rev. Res.},
  volume = {5},
  issue = {2},
  pages = {L022037},
  numpages = {6},
  year = {2023},
  month = may,
  publisher = {American Physical Society},
  doi = {10.1103/PhysRevResearch.5.L022037},
  url = {https://link.aps.org/doi/10.1103/PhysRevResearch.5.L022037}
}

@article{schrieffer_SW_1966,
  title = {Relation between the Anderson and Kondo Hamiltonians},
  author = {Schrieffer, J. R. and Wolff, P. A.},
  journal = {Phys. Rev.},
  volume = {149},
  issue = {2},
  pages = {491--492},
  numpages = {0},
  year = {1966},
  month = sep,
  publisher = {American Physical Society},
  doi = {10.1103/PhysRev.149.491},
  url = {https://link.aps.org/doi/10.1103/PhysRev.149.491}
}

@article{bravyi_SW_2011,
  title     = {{Schrieffer--Wolff} transformation for quantum many-body systems},
  author    = {Bravyi, Sergey and DiVincenzo, David P and Loss, Daniel},
  journal   = {Ann. Phys. (N. Y.)},
  publisher = {Elsevier BV},
  volume    =  {326},
  number    =  {10},
  pages     = {2793--2826},
  month     =  oct,
  year      =  {2011},
}

@article{jansen_mapping_2026,
  title = {Mapping Phase Diagrams of Quantum Spin Systems through Semidefinite-Programming Relaxations},
  author = {Jansen, David and Farina, Donato and Mortimer, Luke and Heightman, Timothy and Leitherer, Andreas and Mujal, Pere and Wang, Jie and Ac\'{\i}n, Antonio},
  journal = {Phys. Rev. Lett.},
  volume = {136},
  issue = {5},
  pages = {050401},
  numpages = {8},
  year = {2026},
  month = {Feb},
  publisher = {American Physical Society},
  doi = {10.1103/j9rb-tnj4},
  url = {https://link.aps.org/doi/10.1103/j9rb-tnj4}
}

@InProceedings{kingma_adam_2014,
	title = {Adam: {A} {Method} for {Stochastic} {Optimization}},
	shorttitle = {Adam},
	url = {https://www.semanticscholar.org/paper/Adam%3A-A-Method-for-Stochastic-Optimization-Kingma-Ba/a6cb366736791bcccc5c8639de5a8f9636bf87e8},
	urldate = {2026-06-15},
	booktitle = {International Conference on Learning Representations},
	author = {Kingma, Diederik P. and Ba, Jimmy},
	month = dec,
	year = {2015},
}

\appendix

\section{Simultaneous Hamiltonian compression in Pauli space}\label{sec:compression_pauli_space}

In this section we discuss an additional step in the CAS set construction for the general case where the eigenbases of $A$ and $B$ are not mutually orthogonal in projective Hilbert space. Equivalently, in this scenario we have that for any partition of $A, B$ of the form
\begin{gather}
    A = \sum_i c_i^A \hat{\sigma}_i = \sum_i h_i^A \ , \\ 
    B = \sum_i  c_i^B \hat{\sigma}_i = \sum_i h_i^B \ ,
\end{gather}
there remain compatible operator contributions satisfying $[h_i^A, h_j^B] = 0$ for some $i, j$.
In particular, we propose an approach to transform our original $A$, $B$ to a basis in which one may find better scoring CAS for our term-removal accessibility criterion. 

Our goal is to find the unitary rotation $U=e^{-i\hat{\tau}}$ that moves us to a basis where a true symmetry $T$ is diagonal, since in such a basis our full term-removal criterion will provide lower-error CAS. Note that, in this basis, the description of $A$ and $B$ is most sparse (assuming $[U, T] \neq 0$), as all basis elements that do not commute with $T$ will be zero or very small. Thus, we can search for unitaries satisfying our criterion by simultaneously minimising the Shannon entropies of the probability distributions induced by the squared Pauli coefficients of $A$ and $B$,
\begin{equation}
    (\vec{p}_{X})_i = \frac{(c^X_i)^2}{\sum_j (c_j^X)^2}  \ .
    \label{eq:pX}
\end{equation}

In order to minimise the entropy of $p_A, p_B$ while still remaining memory-efficient, we will consider $\epsilon$ steps along the unitary manifold ($\epsilon \ll 1)$, such that the infinitesimal transformation
\begin{equation}
    u = e^{-i \epsilon h}
\end{equation}
transforms Hamiltonian $X$ as
\begin{equation}
    X' = u^\dagger X u = X - i \epsilon [h, X] \ .
    \label{eq:var_X}
\end{equation}
In this manner, the functional derivative of $X$ is $\delta X = \frac{d X}{d\epsilon} = -i [h, X]$. 
To identify the gradient of $\mathcal{L} = (S_A)^2 + (S_B)^2$ in the vector space of $h$, our goal is to be able to write the derivative with respect to $\epsilon$ as an inner product between the gradient vector functional and the direction $h$, such that $\frac{d \mathcal{L}}{d\epsilon} = \langle \nabla_h \mathcal{L} , h\rangle$. Since 
\begin{gather}
    \frac{d\mathcal{L}}{d\epsilon} = 2(S_A \frac{d S_A}{d\epsilon} + S_B \frac{d S_B}{d\epsilon})   \ ,
\end{gather}
we can translate this goal directly to the functional $S_X$, so that 
\begin{equation}
    \frac{d S_X}{d\epsilon} = \langle \nabla_h \mathcal{S}_X , h\rangle = \frac{1}{d} \Tr{G_X h}\ .
\end{equation}
Note that the last equality is just the correspondence between the inner product in Pauli space and the inner product in Hilbert space.
We can build the gradient of the Shannon entropy as 
\begin{align}
    \frac{d S_X}{d \epsilon} &= \frac{1}{d} \Tr{ \left(\sum_k \frac{d S_X}{d x_k} \hat{\sigma}_k\right) \left(\sum_j \frac{d x_j}{d \epsilon} \hat{\sigma}_j\right)} = \nonumber \\
    &= \Tr{ \frac{1}{d} \left(\sum_k \frac{d S_X}{d x_k} \hat{\sigma}_k\right) \left(\sum_j \frac{d x_j}{d \epsilon} \hat{\sigma}_j\right)} = \nonumber \\
    &= \Tr{G_X \delta X} \ ,
    \label{eq:shannon_entropy_gradient}
\end{align}
with $G_X = \sum_k \frac{d S_X}{d x_k} \hat{\sigma}_k$ and $X=\sum_k x_k \hat{\sigma}_k$.
By plugging \eqref{eq:var_X} into \eqref{eq:shannon_entropy_gradient}, we find
\begin{align}
    \frac{d S_X}{d \epsilon} &= -i \Tr{G_X [h, X]} = \nonumber \\ 
    &= -i (\Tr{G_XhX} - \Tr{G_X Xh})= \nonumber \\ 
    &= -i \Tr{\ [X, G_X]h  \ } \ .
\end{align} 
In this manner we obtain the expression for the gradient of $\mathcal{L}$ as
\begin{align}
    \frac{d \mathcal{L}}{d \epsilon} &= \langle \nabla_h \mathcal{L}, h\rangle = \nonumber \\ 
    &= -i \Tr{h (2S_A[A, G_A] + 2S_B[B, G_B])} \Rightarrow \nonumber \\
    \nabla_h \mathcal{L} &= -i (2S_A[A, G_A] + 2S_B[B, G_B]) \ .
\end{align}

To find the direction of steepest descent, i.e., the one that minimises $\frac{d \mathcal{L}}{d \epsilon}$, we choose $h^* = -\nabla_h \mathcal{L} = i (2S_A[A, G_A] + 2S_B[B, G_B])$.
With this, we can now iteratively approach a minimum by choosing a small step size $\epsilon$ and iteratively updating $A$ and $B$ according to
\begin{gather}
    A_{k+1} = \frac{1}{\Gamma^A_{k}} (A_k -i\epsilon [h^*_k, A_k]) \ , \\
    B_{k+1} = \frac{1}{\Gamma^B_{k}} (B_k -i\epsilon [h^*_k, B_k]) \ ,
\end{gather}
where the normalisation
\begin{equation}
    \Gamma^X_{k} = \frac{1}{||X_k - i \epsilon [h^*_k, X_k]||}
\end{equation}
at each step compensates for some of the error of truncating the unitary at first order by keeping the norm fixed. From here, this simple gradient descent can be easily upgraded to more sophisticated approaches~\cite{kingma_adam_2014}. 

\section{Integral cost function}\label{sec:integral_cost}

This Appendix discusses the results of the CAS reduction when we consider minimising the average $||\nu||_2$ instead of its maximum. Namely, instead of Eq.~\eqref{eq:nu_max_cost} we minimise
\begin{equation}
	\varepsilon^{\text{int}} = \mathcal{C}_{\text{int}}(\nu) = \int_0^1||\nu(s)||_2 ds  \ .
    \label{eq:nu_integral_cost}
\end{equation}
Due to the example we use to discuss the CAS reduction resulting from this cost, we recommend going through the numerical results section beforehand in order to fully appreciate the contrast in the results.

In general, by minimising the average off-diagonal cost we find that the prediction of features like the location of minima in the spectral gap is worse, although the difference is not very noticeable for small system sizes. Nonetheless, for the $N=101$ frustrated Ising ring, minimising Eq.~\eqref{eq:nu_integral_cost} captures the spectral behaviour of both the continuous and the first order phase transition without the addition of any biasing weights, as shown in Fig.~\ref{fig:frustrated_ising_ring_N101_integral_cost}. We comment that, similarly to the unbiased case minimising~\cref{eq:nu_max_cost}, we only find type I CAS apart from the initial exact case II symmetry. Also similar to the unbiased case of the main text, the degeneracy of the first excited pseudo-manifold begins to break shortly before the true critical point $s_c$, a feature visible in the decrease of $\chi_{k>0, j>0}$ in Fig.~\ref{fig:frustrated_ising_ring_N101_integral_cost}a. In addition, the pGS joins the large strongly hybridised cluster shortly before $s_c$ as well, as shown in the inset within the same plot. The behaviour of the off-diagonals is also significantly altered around $s_c$, as shown in Fig.~\ref{fig:frustrated_ising_ring_N101_integral_cost}b, a feature that is also found in the max-cost unbiased case of the main text but that is enhanced here. However, the predicted location of the critical point as per the shape of the pseudo-spectrum in Fig.~\ref{fig:frustrated_ising_ring_N101_integral_cost}c, namely the first crossing of the pGS, is rather far: $s_c^{\text{pred}} = 0.6332 \pm 0.0025$, resulting in a 23\% of relative error. The prediction of the location of the first order transition is also considerably worse than for the max-cost; in this case $s_{\min}^{\text{pred}} = 0.7688 \pm 0.0025$, which constitutes a relative error of 14\%. We note that the hybridisation remains quite large between all pairs of pseudo-levels for $s_c \lesssim s \lesssim s_{\min}$ for this CAS reduction, which indeed implies a large potential displacement of the relevant features that the pseudo-spectrum shows in this region. The uncertainty present in the bounds is also somewhat larger than in the max-cost case, which can be explained by the fact that the found $||\nu||_2$ is slightly larger, and just as for the max-cost, $L=20$ provides bounds that are as faithful as if we were to consider the full first excited pseudo-manifold for $s\lesssim s_c$.

All in all, despite the arguably better qualitative agreement, the overall higher errors found by minimising the average off-diagonal 2-norm led us to search for the CAS by minimising the max-cost in the main text.

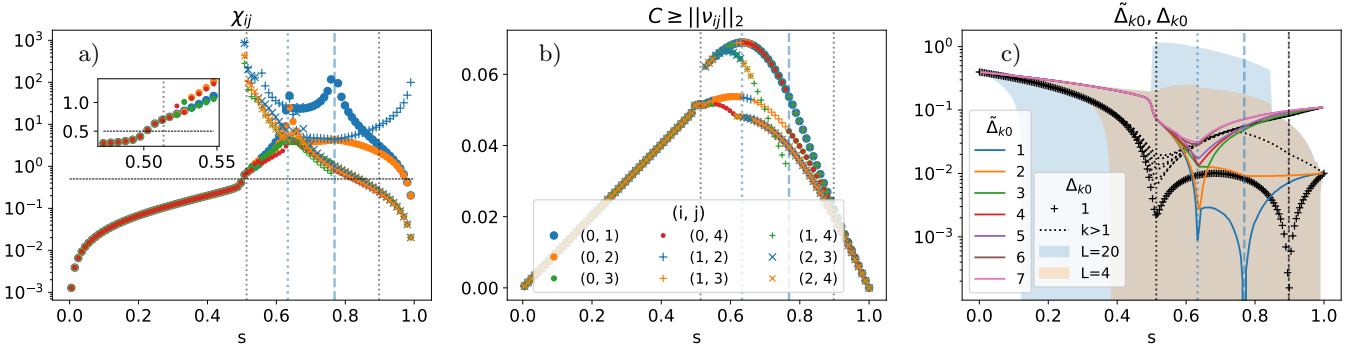
\begin{figure*}
\centering
    \begin{tikzpicture}
      \node[anchor=south west, inner sep=0] (img) at (0,0)
        {\includegraphics[width=\linewidth]{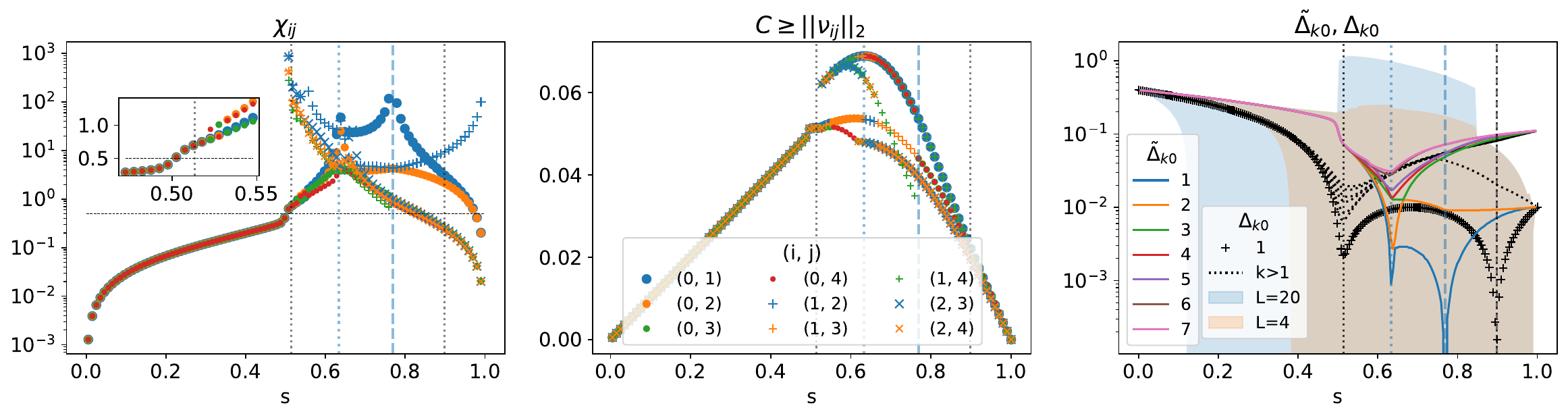} };
      \begin{scope}[x={(img.south east)}, y={(img.north west)}]
        \node[] at (0.07,0.83) {a)};
        \node[] at (0.41,0.83) {b)};
        \node[] at (0.75,0.83) {c)};
      \end{scope}
    \end{tikzpicture}
  \caption{Simulation of the $N=101$ frustrated Ising ring where the optimiser considered minimising the integral of $C$ (Eq.~\eqref{eq:nu_integral_cost}), without biasing weights. The vertical black lines mark the true transition points $s_c$ (dotted) and $s_{\min}$ (dashed), while their blue counterparts mark the locations predicted by the pseudo-spectrum. Figure a) shows the hybridisation between all pairs within the lowest four pseudo-levels, with a close-up around the true $s_c$ shown in the inset. Plot b) shows the upper bound $C$ on the 2-norm of the off-diagonals (see Eq.~\eqref{eq:C_upper_bound_nu2}) that connects the different pairs of the four lowest-lying pseudo-eigenstates. Since all the nontrivial CAS symmetries found were of type I, in this case we actually have $C = ||\nu||_2$. Plot c) presents the first seven energy separations with the true GS (black) and with the pGS (colours), where we set to zero values below $10^{-4}$ to aid visibility. The shaded regions indicate the gap bounds calculated considering the lowest $L$ pseudo-levels. We highlight that the occasional discontinuities in a) and b) are an artifact of pseudo-level degeneracy.}
  \label{fig:frustrated_ising_ring_N101_integral_cost}
\end{figure*}

\section{Proof of bounds on the spectral gap}\label{sec:proof_bounds_gap}

The derivations below follow the three hybridisation regimes introduced in the bounds discussion in the Results: low-lying pseudo-levels weakly hybridised with the rest of the spectrum, a weakly hybridised pGS coupled to a strongly hybridised excited-state cluster, and a strongly hybridised cluster containing the pGS itself.


\subsection{Proof of~\cref{eq:bounds_gap}}

We now proceed to derive the bounds in the base case of weak hybridisation where only two pseudo-levels are relevant to the gap.
Thus, we have $\chi_{0k} < 1/2$ and $\chi_{1k} < 1/2$ for any $k> 1$. 
In this scenario, we can bound the error $|\Delta(s) - \tilde{\Delta}(s)|$ via the straight-forward application of Weyl's inequalities, which are a standard tool in linear algebra (see, for example, ~\cite{bhatia_matrix_1997} for a more in-depth description). Applied to our case, Weyl's inequality states that for a decomposition of a Hamiltonian $H$ into a diagonal part $\tilde{H}$, where the pseudo-eigenvalues lie, and an off-diagonal part $\nu$, such that $H=\tilde{H} + \nu$, we have
\begin{equation}
    |\lambda_k(H) - \lambda_k(\tilde{H})|  \leq  ||\nu||_2 \ ,
    \label{eq:Weyl_inequality_evals}
\end{equation}
where $\lambda_k(X)$ denotes the $k$-th eigenvalue of $X$ and $||\nu||_2 = \lambda_{\max}(\sqrt{\nu^\dagger\nu})$ is the 2-norm. Consequently, we can conclude that
\begin{gather}
    |\lambda_j(H) - \lambda_j(\tilde{H}) - \lambda_k(H)  + \lambda_k(\tilde{H})| \leq 2 ||\nu_{kj}||_2 \Rightarrow \nonumber\\
    |\Delta_{kj} - \tilde{\Delta}_{kj}|  \leq  2||\nu_{kj}||_2 \ ,
    \label{eq:Weyl_inequality_gap}
\end{gather}
where $\nu_{kj}$ is the off-diagonal block that connects pseudo-eigenlevels $k, j$, $\Delta_{kj} = \lambda_k(H) - \lambda_j(H)$ and $\tilde{\Delta}_{kj} = \lambda_k(\tilde{H}) - \lambda_j(\tilde{H})$ for $k > j$.

Furthermore, whenever $||\nu||_2 < \tilde{\Delta}/2$ we can introduce a tighter bound on the true eigenvalues and, thus, on the spectral gap. Notice that, under such condition, Eq.~\eqref{eq:Weyl_inequality_evals} guarantees that the true eigenvalues won't exchange their ordering with respect to the ordering of their associated pseudo-eigenvalues. Let us consider the block decomposition
\begin{gather}
    H = \tilde{H} + \nu = \begin{pmatrix}
        \tilde{H}_+ & 0 \\
        0 & \tilde{H}_-
    \end{pmatrix} +  
    \begin{pmatrix}
        0 & \nu_{\pm} \\
        \nu^\dagger_{\pm} & 0
    \end{pmatrix} \ .
\end{gather}
Furthermore, let us assume that $|\lambda_i(\tilde{H}_+) - \lambda_j(\tilde{H}_-)| > ||\nu||_2$ $\forall i, j$. Now, to determine the eigenvalues of $H$ we write the eigenvalue equation
\begin{gather}
    \begin{pmatrix}
        \tilde{H}_+ - \lambda I & \nu_{\pm} \\
        \nu_{\pm}^\dagger & \tilde{H}_- - \lambda I
    \end{pmatrix}
    \begin{pmatrix}
        x \\ y
    \end{pmatrix}
    = \begin{pmatrix}
        0 \\ 0
    \end{pmatrix} \ .
\end{gather}
Thanks to the Weyl inequality we know that $\tilde{H}_- - \lambda I$ is non-singular for the part of the spectrum that originates from $\tilde{H}_+$. Thus, its inverse is well defined and we can solve the system of equations as
\begin{gather}
    y = -(\tilde{H}_- - \lambda I)^{-1}\nu_{\pm}^\dagger x \ ,\\
    \left(\tilde{H}_+ - \lambda I) - \nu_{\pm} (\tilde{H}_- - \lambda I)^{-1}\nu_{\pm}^\dagger\right) x = 0  \ .
\end{gather}
As a result, the sought-after eigenvalue becomes an eigenvalue of the (nonlinear) effective Hamiltonian
\begin{equation}
    \tilde{H}_+^{\text{eff}} = \tilde{H}_+ - \nu_{\pm} (\tilde{H}_- - \lambda I)^{-1}\nu_{\pm}^\dagger \ .
    \label{eq:schur_complement}
\end{equation}
Eq.~\eqref{eq:schur_complement} is precisely the Schur complement of $\tilde{H}_- - \lambda I$ in $H-\lambda I$ (see, for example, ~\cite{zhang_schur_2005}), which 
is also known as the Feshbach effective Hamiltonian~\cite{griesemer_smooth_2008}.
Now we can apply Weyl's inequality to the effective Hamiltonian~\eqref{eq:schur_complement}, considering the diagonalising basis of $\tilde{H}_+$ and knowing that $\nu_{\pm} (\tilde{H}_- - \lambda I)^{-1}\nu_{\pm}^\dagger$ is fully off-diagonal with respect to it. The next step is to bound the 2-norm of this off-diagonal, which we can readily do with the help of the well-established bound on the norm of the resolvent $(X - \lambda I)^{-1}$ for any normal operator $X$, $||(X - \lambda I)^{-1}|| \leq \frac{1}{\text{dist}(\lambda, \text{spec}(X))}$\cite{kato_perturbation_1995}. In addition, since $\text{dist}(\lambda, \text{spec}(X)) \geq \min_{i, j} [\lambda_i (\tilde{H}_+) -  \lambda_j (\tilde{H}_-)] - ||\nu||_2$ we have
\begin{gather}
    || \nu_{\pm} (\tilde{H}_- - \lambda I)^{-1}\nu_{\pm}^\dagger|| \leq ||\nu_{\pm}||^2 ||(\tilde{H}_- - \lambda I)^{-1}|| \Rightarrow  \nonumber \\
    || \nu_{\pm} (\tilde{H}_- - \lambda I)^{-1}\nu_{\pm}^\dagger|| \leq \frac{(||\nu||_2)^2}{\min_{i, j} [\lambda_i (\tilde{H}_+) -  \lambda_j (\tilde{H}_-)] - ||\nu||_2}
\end{gather}
Thus, taking $\tilde{\Delta}_{\text{min}}^{\pm} := \min_{i, j} [\lambda_i (\tilde{H}_+) -  \lambda_j (\tilde{H}_-)]$, the shift on the eigenvalues of $H$ that are stably perturbed from $\tilde{H}_+$ is
\begin{equation}
    |\lambda_k(H) - \lambda_k(\tilde{H})| \leq \frac{\left(||\nu||_2\right)^2}{\tilde{\Delta}_{\text{min}}^{\pm} - ||\nu||_2} \ .
\end{equation}
Going over the same procedure focusing on the eigenvalues that are connected to the $-$ sector provides the same bound, and thus we obtain the bounds presented in the main text:
\begin{gather}
    |\Delta_{kj} - \tilde{\Delta}_{kj}|  \leq 
    \left\{
    \begin{array}{lr}
        \frac{2 \left(||\nu_{kj}||_2\right)^2}{\tilde{\Delta}_{kj} - ||\nu_{kj}||_2} & \text{if } ||\nu_{kj}||_2 < \tilde{\Delta}_{kj} / 2 \\
        2 ||\nu_{kj}||_2 & \text{otherwise}
    \end{array}
    \right.
\end{gather}


\subsection{Proof of~\cref{eq:lower_bounds_gap_deg1exc,eq:upper_bounds_gap_deg1exc}}

We now consider the scenario corresponding to the excited-state cluster regime, where $0 \notin \kappa$ with $\kappa$ denoting the strongly hybridised connected component containing the first pseudo-level.


The true interactions within the cluster are described by the matrix $V_{\text{int}}$,
\begin{equation}
    (V_{\text{int}})_{ij} = \bra{\mu_i} \nu \ket{\mu_j} , \quad i, j \in \kappa \ .
    \label{eq:V_int}
\end{equation}
Since the entries of $V_{\text{int}}$ are not efficiently accessible, we replace them by the available upper bounds $||\nu_{ij}||_2$. This leads to the effective interaction matrix introduced in the Results section,
\begin{gather}
    (v_{\text{int}})_{ij} = \left\{ \begin{array}{cc}
        ||\nu_{ij}||_2 & i \neq j \\
        0 & i =j
    \end{array} \right. \ .
\end{gather}
Note that, indeed, $||V_{\text{int}}||_2 \leq || \ | V_{\text{int}} | \ ||_2 \leq ||v_{\text{int}}||_2$.
We can then define the corrected lower bound for the first excited pseudo-level as
\begin{gather}
    \mu_{1, l}^\prime = \mu_1 - ||v_{\text{int}}||_2 \ .
\end{gather}
This correction yields the modified pseudo-gap estimate introduced in the Results section,
\begin{equation}
    \tilde{\Delta}_{10, l}^\prime = \tilde{\Delta}_{10} - ||v_{\text{int}}||_2 \ .
\end{equation}
Applying the variational principle onto the cluster sub-block one can see that an upper bound is given by the lowest pseudo-level of the cluster itself, such that $\mu_{1, u}^\prime = \mu_1$.

On the other hand, the interaction between the pGS and any state living in the cluster is upper bounded by $||v_{\text{eff}}||_2$,
\begin{equation}
    ||v_{\text{eff}}||_2 = \sqrt{\sum_{k \in \kappa} \left(||\nu_{0k}||_2\right)^2} \ .
\end{equation}
For the upper bound on $\Delta$, however, since $\mu^\prime_{1, u}$ assumes no hybridisation of the first pseudo-level with the rest of the cluster, the off-diagonal that should be taken into account is $||\nu_{01}||_2$ alone. Thus, the upper bound is analogous to the unclustered case.
Putting all of this together we arrive at the bounds stated in the Results section,
\begin{gather}
    \Delta_{10} \leq \tilde{\Delta}_{10} + \left\{
    \begin{array}{ll}
         \frac{2 \left(||\nu_{01}||_2\right)^2}{\tilde{\Delta}_{10} - ||\nu_{01}||_2} & \text{if } ||\nu_{01}||_2 < \tilde{\Delta}_{10} / 2 \\
        2 ||\nu_{01}||_2 & \text{otherwise}
    \end{array}
    \right.
    \\
    \Delta_{10} \geq \tilde{\Delta}^\prime_{10, l} - \left\{
    \begin{array}{ll}
         \frac{2 \left(||\nu_{\text{eff}}||_2\right)^2}{\tilde{\Delta}^\prime_{10, l} - ||\nu_{\text{eff}}||_2} & \text{if } ||\nu_{\text{eff}}||_2 < \tilde{\Delta}^\prime_{10, l} / 2 \\
        2 ||\nu_{\text{eff}}||_2 & \text{otherwise}
    \end{array}
    \right.
\end{gather}

As an operational remark, we highlight that the quantity $||v_{\text{int}}||_2$ can be computed via exact diagonalisation efficiently if the cluster's dimension is polynomial in system size. 
Note that this is the only relevant regime in practice, since we are limited to tracking a polynomial number of branches. We highlight that one can also choose to keep the analytical $s$-dependence by considering upper bounds on $||v_{\text{int}}||_2$ instead of the value resulting from exact diagonalisation. A possible bound we can compute analytically is the Frobenius norm,
\begin{equation}
    ||v_{\text{int}}||_F = \sqrt{\Tr{v_{\text{int}}^\dagger v_{\text{int}} }} = \sqrt{\sum_{i, j \in \kappa} (||\nu_{ij}||_2)^2}  ,
\end{equation}
which always upper-bounds the 2-norm. Alternatively, if it is more advantageous one can consider the bound $C_G$ on the 2-norm given by Gershgorin's circle theorem,
\begin{equation}
    C_G = \max_j \sum_i (v_{\text{int}})_{ji} = \max_{j\in \kappa} \sum_{i\in \kappa} ||\nu_{ij}||_2 \ .
\end{equation}
Thus, we could consider $\min(C_G, ||v_{\text{int}}||_F) \geq ||v_{\text{int}}||_2 $ as well for constructing $\mu_{1, l}^\prime, \mu_{1, u}^\prime$ in order to keep the treatment fully analytical.



\subsection{Proof of~\cref{eq:bounds_gap_clusterGS}}

We now consider that the lowest-lying hybridised cluster $\kappa$ contains the pGS, i.e., $0 \in \kappa$. In this case, the lower bound is automatically fixed to 0, since we have no guarantee that the gap doesn't close. For the upper bound we could still apply Weyl directly, but since $||v_{\text{int}}||_2$ is extensive in cluster size this will provide very unrealistic estimates. Indeed, one can notice that if the pGS and first excited pseudo-level maximally repelled each other, the gap would be given by the pGS and whichever higher excited pseudo-state was left lowest after the correction. We can account for this intuition with the help of the min-max theorem, which will provide tighter results than Weyl alone. According to min-max, we can bound $\lambda_1$ as
\begin{equation}
    \lambda_1 \leq \min_{
        S=\text{span}\left\{\ket{\mu_0}, \ket{\mu_j} \right\}} \ 
        \max_{\ket{\psi} \in S} \  \bra{\psi} H\ket{\psi} \ ,
\end{equation}
%
The maximum possible eigenvalue of each such 2-dimensional sub-block of $H$ is 
\begin{align}
    &\max_{\ket{\psi} \in \text{span}\left\{\ket{\mu_0}, \ket{\mu_j}\right\}} \  \bra{\psi} H\ket{\psi} = \nonumber \\
    &= \frac{\mu_0 + \mu_j}{2} + \sqrt{\left( \frac{\tilde{\Delta}_{j0}}{2} \right)^2 + (||\nu_{0j}||_2)^2} \ . 
\end{align}
The upper bound on the spectral gap thus becomes
\begin{equation}
    \lambda_1 - \lambda_0 \leq \min_j \frac{\mu_0 + \mu_j}{2} + \sqrt{\left( \frac{\tilde{\Delta}_{j0}}{2} \right)^2 + (||\nu_{0j}||_2)^2} - \lambda_0 \ ,
\end{equation}
which recalling the Weyl lower bound on the pGS,
\begin{equation}
    \lambda_0 \geq \mu_0 - ||v_{\text{int}}||_2
\end{equation}
becomes
\begin{equation}
    \lambda_1 - \lambda_0 \leq \min_{j\in \kappa} \frac{\mu_j - \mu_0}{2} + \sqrt{\left( \frac{\tilde{\Delta}_{j0}}{2} \right)^2 + (||\nu_{0j}||_2)^2} + ||v_{\text{int}}||_2 \ .
\end{equation}
Thus, we are left with the bound provided in the Results section,
\begin{equation}
    0 \leq  \Delta_{10} \leq 
    ||v_{\text{int}}||_2 +
    \min_{j\in \kappa} \frac{\tilde{\Delta}_{j0}}{2} + \sqrt{\left( \frac{\tilde{\Delta}_{j0}}{2} \right)^2 + (||\nu_{0j}||_2)^2} \ . 
\end{equation}
Despite being tighter than Weyl alone, as it still partially depends on it, the latter upper bound remains substantially loose for large clusters.

\section{Violation and restoration of bounds on the spectral gap}\label{sec:restoration_bounds}

As mentioned in the main text, if the discarded off-diagonal $\nu$ is too great we may run into the situation in which the bounds are not necessarily satisfied any more. Specifically, this error can cause some pseudo-levels to shift very far from the true eigenlevels they are closest to, potentially reordering them significantly (i.e., taking them outside what would be considered a hybridised cluster) and thus producing the illusion of a larger pseudo-gap and less low-lying pseudo-levels. This shift becomes important when the eigendirections of $\nu$ become parallel to the true eigenvectors. Within the CAS Hamiltonian and CAS families under consideration in Section~\ref{sec:TFIM_example}, we identified that this situation sometimes arises in the scenario where there is a case IV reduction followed by some case II. A case IV (for a homogeneous initial Hamiltonian $A$) is already unlikely due to our normalisation, but we observe it is the type of the first CAS reduction for random Ising instances where the gap remains large throughout the interpolation. Note that case IV modifies the local fields of the projected Hamiltonians in an $s$-independent manner (see Eq.~\eqref{eq:Bpm_caseIV}), which can then imply a large $\nu$ contribution around $s=1$ if we have a case II (see Eq.~\eqref{eq:MB_caseII}) coming from such a previous reduction step, even though the off-diagonal $\nu_{IV} = (1-s)M_A$ is 0 at $s=1$ as per Eq.~\eqref{eq:MA_caseIV}.

This violation can be corrected by explicitly accounting for the induced shift of the branches affected by the case IV - case II combination by considering the lower bound on the branch's pseudo-level, given by accounting for $\nu_{II}$ as diagonal contributions to the pseudo-energies near $s=1$ (where, in fact, they are). The bounds calculated with these corrected pseudo-level estimates are then enclosing the true gap once again. An example of this is shown in Fig.~\ref{fig:violation_restoration_bounds}, where an instance that presents such a violation of the bounds is shown, along with the amended ones near $s=1$. 

\begin{figure}
    \centering
    \includegraphics[width=\linewidth]{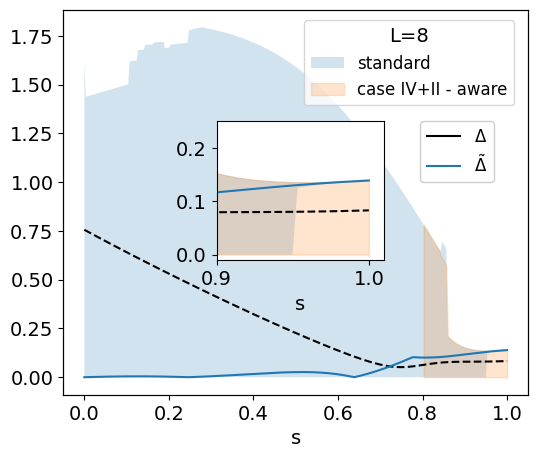}
    \caption[Violation and restoration of gap bounds in an edge case.]{True spectral gap (black), pseudo-gap (solid, blue) and gap bounds for a random instance of $N=7$ with a large $\nu$ due to the combination of type IV and type II CAS in some branches that correspond to low-lying states. The shaded blue region corresponds to the bounds resulting from the straight-forward application of~\cref{eq:bounds_gap,eq:lower_bounds_gap_deg1exc,eq:upper_bounds_gap_deg1exc,eq:bounds_gap_clusterGS}, which fail around $s\approx 0.95$. The shaded orange region corresponds to the corrected bounds taking into account the shift of the IV+II CAS-type branches by considering the $\nu_{II}$ contributions as diagonal for $s>0.8$.}
    \label{fig:violation_restoration_bounds}
\end{figure}

\section{CAS reductions for interpolations between the Ising model and a transverse field}\label{sec:CAS_reductions}
Here we present the concrete expressions of the CAS reduction types developed for the study of the interpolation between the transverse field Hamiltonian and an Ising model. 


\subsection{Case I: remove all couplings to a single qubit}\label{sec:caseI}

If we disconnect the $k$-th qubit from the rest, the resulting symmetry is the operator along the remaining local field in the now isolated qubit. 
The explicit shape of the CAS is thus 
\begin{align}
    T_{\text{I}} = & (\cos{\varphi} X_k + \sin{\varphi} Z_k) = \\
    =& \frac{(1-s)h^x_k}{\sqrt{((1-s)h^x_k)^2 + (s h^z_k)^2}}X_k + \nonumber \\
    &+ \frac{sh^z_k}{\sqrt{((1-s)h^x_k)^2 + (s h^z_k)^2}} Z_k
    \label{eq:T_I} 
\end{align}
where
\begin{gather}
    \cos{\varphi} = \frac{(1-s)h^x_k}{\eta} \quad , \quad
    \sin{\varphi} = \frac{s h^z_k}{\eta} 
    \label{eq:cos_sin_varphi}\\
    \text{and }\ \eta = \sqrt{((1-s)h^x_k)^2 + (s h^z_k)^2} \ .
\end{gather}
Note that $T_{\text{I}}$ is directly proportional to $h^z_k$, so its sign is what determines whether the $-1$ sector corresponds to a spin up or to a spin down with respect to the instantaneous field.

In order to find the projected sectors and off-diagonal terms, we must first rotate to the CAS basis by means of the relevant unitary $U$. The transformed $Z_{i\neq k}Z_k$ terms become
\begin{gather}
    U^\dagger Z_{i\neq k}Z_k U = \cos{\varphi} Z_{i\neq k} \tilde{X}_k + \sin{\varphi} Z_{i\neq k} \tilde{Z}_k \ ,
\end{gather}
where the tilde indicates that the operator is in the CAS basis.
Thus, the off-diagonal parts with respect to the CAS basis are
\begin{gather}
    M_A = 0 \ ,\\
    M_B = \sum_{i\neq k} J_{ik} \frac{(1-s)h^x_k}{\eta} Z_i \tilde{X}_k \ .
\end{gather}

In order to find the identity contributions to the sub-sectors of $A$, $\mathcal{I}_A$, we need to consider that $U^\dagger X U= \cos{\varphi} \tilde{Z} - \sin{\varphi} \tilde{X}$. With this, we find that the identity contributions on each sub-sector are
\begin{gather}
    \mathcal{I}_A = h^x_k \cos{\varphi} = \frac{ (1-s) (h^x_k)^2}{\eta} \ ,
    \label{eq:IA_caseI}
    \\
    \mathcal{I}_B = h^z_k \sin{\varphi} = \frac{s (h^z_k)^2}{\eta}\ .
    \label{eq:IB_caseI}
\end{gather}
Then, the reduced Hamiltonians in each sub-sector ($\prom{\tilde{Z}_k} = \pm 1$) are
\begin{gather}
    A_{\pm} = \sum_{i\neq k} h^x_i X_i \pm \mathcal{I}_A \ ,
    \label{eq:Apm_case1} \\
    B_{\pm} = \sum_{i\neq k} \left(h^z_i \pm J_{ki} \frac{s h^z_k}{\eta} \right) Z_i + \sum_{\substack{i, j \neq k\\ j<i}} J_{ij} Z_i Z_j \pm \mathcal{I}_B \ .
    \label{eq:Bpm_case1}
\end{gather}
In this manner, the full reduced Hamiltonian on each sub-sector becomes
\begin{equation}
    H_{\pm}(s) = (1-s) A_{\pm} + s B_{\pm}\ .
\end{equation}


\subsection{Case II: remove all $z$-fields}\label{sec:caseII}

When coupling energies dominate, we can have the case where removing all single-qubit terms in $z$ is the cheapest option with respect to the cost $\varepsilon$ in Eq.~\eqref{eq:nu_max_cost}, which then endows $H(s)$ with a symmetry $T_{\text{II}}$ in the projected CAS basis
\begin{equation}
    T_{\text{II}}=\prod_{i=0}^{N-1} X_i 
    \label{eq:T_II}
\end{equation}
that renders off-diagonal contributions
\begin{gather}
    M_A = 0 \ ,\\
    M_B = \sum_{i=0}^{N-1} h^z_i Z_i \ .
    \label{eq:MB_caseII}
\end{gather}
Since in this case the CAS is independent of $s$, we can perform a static unitary transformation to obtain the reduced Hamiltonians in their respective sectors. This basis change is not unique but rather embodied by a family of transformations (see~\cite{fujii_eigenvalue-invariant_2023,palacios_scalable_2025})
This transformation maps all variables to the parity with respect to a chosen qubit, which we index by $k$, such that:
\begin{gather*}
    Z_0 Z_k \to \tilde{Z}_0 \\
    \vdots \\
    Z_{k-1} Z_k \to \tilde{Z}_{k-1} \\
    Z_{k+1} Z_k \to \tilde{Z}_k \\
    \vdots \\
    Z_{N-1} Z_k \to \tilde{Z}_{N-2}
\end{gather*}
where the tilde indicates the transformed basis. This transformation can be found by thinking in terms of a quantum circuit, as mentioned in \cite{fujii_eigenvalue-invariant_2023}, since the CNOT operation enacts an XOR on the target qubit. Consequently, the expression of $B_{\pm}$ is straight-forward, and only the derivation of $A_{\pm}$ is missing. If we consider the full unitary as $U = H^{\otimes N} \prod_{i\neq k} CNOT_{i\to k} H^{\otimes N}$, where $H^{\otimes N}$ is the product of Hadamard gates acting on every qubit, we find that
\begin{align}
    U^\dagger X_j U = \begin{cases}
        \tilde{X}_j \quad &\text{if }j\neq k\\
        \prod_{i\neq k}^{N-2} \tilde{X}_i \tilde{X}_k \quad &\text{if }j= k
    \end{cases} \ .
\end{align}
Thus, the reduced Hamiltonians on each sub-sector are
\begin{gather}
    A_{\pm} = \sum_{i\neq k} h^x_i X_i \pm h^x_k \prod_{i\neq k}^{N-2} X_i \ ,
    \label{eq:Apm_case2}\\
    B_{\pm} = \sum_{i\neq k} J_{ik} Z_i + \sum_{\substack{j<i \\ i, j \neq k}} J_{ij} Z_i Z_j \ ,
    \label{eq:Bpm_case2}
\end{gather}
as $\prom{\tilde{X}_k}_{\pm} = \pm 1$ in the CAS sub-blocks. We note that, in this case, $\mathcal{I}_A = \mathcal{I}_B = 0$.
We highlight that, despite the appearance of a many-body term on the reduced problems, the possible types of CAS under consideration do not change; we merely need to keep track of these many-body terms because now they may contribute to the diagonal and/or off-diagonal parts as well in subsequent CAS rounds (see Appendix~\ref{sec:many_body_terms_treatment} for details).

We highlight that, despite the fact that the associated cost~\eqref{eq:nu_max_cost} of a case II reduction is independent of the choice of ``pivot" qubit $k$, the resulting projected Hamiltonians are not equal, thus leading to different costs in the next CAS reductions. For this reason, in order to choose the optimal $k$ we adopt a look-ahead strategy; if case II arises at recursion order $r$, we compute the $n_{\text{eff}} = N - r$ alternative projections, evaluate the optimal CAS in all of them and choose the $k$ that produced the minimal $\mathcal{C}(\nu^{(r+1)})$. Since the search over all possible CAS is polynomial in $n_{\text{eff}}$, this look-ahead strategy remains feasible in polynomial time as well. Note that, in principle, this look-ahead strategy could be extended further than one step towards the future. However, assuming homogeneity in the $|\bra{\mu_k} \nu \ket{\mu_j}|$ in the average case, since the dimension of $\nu^{(r)}$ decreases exponentially with the order $r$ we can consider that lower orders of $r$ will be more important than subsequent ones. For this reason, we restrict the algorithm to searching through the immediately upcoming iteration only.


\subsection{Case III: remove some $z$-fields and some couplings}\label{sec:caseIII}

As an extension of case II, we can consider symmetries of the type $T_{\text{III}}$, with shape
\begin{equation}
    T_{\text{III}} = \prod_{i\neq\mathcal{M}} X_i \ ,
    \label{eq:T_III} 
\end{equation}
where the excluded variables belong to the set $\mathcal{M}=\{p_k\}_{k=0}^{m-1}$ for $1 < m < N-2$.
This CAS type corresponds to the case in which we can split the problem graph into two complementary sets of variables which are weakly connected to each other, $\mathcal{M}$ and $\mathcal{R}$, with $|\mathcal{M}| = M$ and $|\mathcal{R}| = R = n_{\text{eff}}-M$. We recall that $n_{\text{eff}}$ is the current dimension of the block. Furthermore, in this scenario set $\mathcal{R}$ has small local fields overall, and thus an approximate $\mathbb{Z}_2$ symmetry within its support. To verify that the reduction indeed produces two invariant sectors of equal dimension in the full current block, it is useful to inspect the corresponding dynamical Lie algebras: in the CAS projection we have $\text{DLA}(\mathcal{M})= \mathfrak{su}(2^M)$ and $\text{DLA}(\mathcal{R})= \mathfrak{su}(2^{R-1}) \oplus \mathfrak{su}(2^{R-1})$, such that 
\begin{align}
    \text{DLA}_{tot} &= 
    \text{DLA}(\mathcal{M} , \mathcal{R}_+) \oplus \text{DLA}(\mathcal{M} , \mathcal{R}_-) 
    =\nonumber\\
    &= \mathfrak{su}(2^M \cdot 2^{R-1})  \oplus  \mathfrak{su}(2^M \cdot 2^{R-1})  =\nonumber\\
    &= \mathfrak{su}(2^{n_{\text{eff}}-1}) \oplus \mathfrak{su}(2^{n_{\text{eff}}-1}) \ .
\end{align}

The resulting $M_A, M_B$ in this scenario are
\begin{gather}
    M_A = 0 \ ,\\
    M_B = \sum_{i\in \mathcal{R}} h^z_i Z_i + \sum_{i\in \mathcal{R}} \sum_{j \in \mathcal{M}} J_{ij} Z_i Z_j \ .
    \label{eq:Mb_caseIII}
\end{gather}
Since only the set $\mathcal{R}$ is being altered, the part of the Hamiltonian acting exclusively in $\mathcal{M}$ remains the same and we can just add it to the transformed part, described in \cref{eq:Apm_case2,eq:Bpm_case2}. This renders projected Hamiltonians of the shape
\begin{gather}
    A_{\pm} = \sum_{i\neq k} h^x_i X_i \pm h^x_k \prod_{\substack{i\in \mathcal{R} \\ i\neq k}} X_i  \ ,
    \label{eq:Apm_case3}\\
    B_{\pm} = \sum_{\substack{i\in \mathcal{R} \\ i\neq k}} J_{ik} Z_i + \sum_{\substack{i, j \in \mathcal{R} \\ j<i \\ i, j \neq k}} J_{ij} Z_i Z_j + \sum_{i \notin \mathcal{R}} h_i Z_i + \sum_{\substack{i, j \notin \mathcal{R} \\ j<i}} J_{ij} Z_i Z_j \ .
    \label{eq:Bpm_case3}
\end{gather}
Note that a qubit $k$ must still be chosen within the set $\mathcal{R}$, as in case II. The criterion we propose for this selection is also analogous to case II; the optimisation of the cost of the next CAS splitting.


The minimisation of the cost associated with $M_B$ as defined in Eq.~\eqref{eq:Mb_caseIII} to find the best subset $\mathcal{R}$ may seem a bit of a daunting task, especially when one considers that the local fields $h^z_i$ may be $s$-dependent due to a recursion history containing case I CAS. Indeed, even with static $h_i^z$, the graph cut problem that minimising $M_B$ represents is NP-hard if the $J_{ij}$ can have arbitrary signs. However, we can always minimise the upper bound $|| M_B^\prime ||_2 \geq ||M_B||_2$, where
\begin{equation}
    M_B^\prime = \sum_{i\in \mathcal{R}} |h^z_i| Z_i + \sum_{i\in \mathcal{R}} \sum_{j \in \mathcal{M}} |J_{ij}| Z_i Z_j \ .
    \label{eq:Mb_prime_caseIII}
\end{equation}
Note that, for $s$-dependent $h_i^z$, we can define the problem taking the maximum $\max_s |h_i^z(s)|$ or the average $\overline{h}_i^z=\int_0^1 |h_i^z(s)| ds$ as a proxy.
The minimisation of Eq.~\eqref{eq:Mb_prime_caseIII} for static $h_i^z$ is an extension of the minimum cut problem, which can be solved exactly in polynomial time via max-flow algorithms~\cite{ahuja_network_1993}. This efficiency arises from~\eqref{eq:Mb_prime_caseIII} being a submodular function, which is essentially the discrete analogue of a convex function.
Max-flow algorithms are well-established in the literature, and their computational complexity can vary with the structure of the graph, but generic, state-of the art implementations are $O(N^3)$~\cite{orlin_max_2013} (more precisely, $O(N|E|)$ for a graph with $N$ vertices and $E$ edges).
In conclusion, the identification of an optimal subset $\mathcal{R}$ can be done efficiently, and then the only issue that remains is the choice of $k$ among these qubits. This is precisely the same issue we had to solve for case II, and just as before we can tackle it efficiently by considering the choice that minimises the upcoming CAS error.



\subsection{Case IV: remove a single x-field on some qubit}\label{sec:caseIV}

The last way to induce a $\mathbb{Z}_2$ symmetry is a rather trivial one: if we remove the $x$-component to some qubit $k$, we will eliminate its dynamics. Thus, the CAS $T_{\text{IV}}$ is simply
\begin{equation}
    T_{\text{IV}} = Z_k \ ,
    \label{eq:T_IV}
\end{equation}
which induces off-diagonal parts
\begin{gather}
    M_A = h^x_k X_k \ ,
    \label{eq:MA_caseIV} \\
    M_B = 0 \ ,
\end{gather}
and corresponding identity contributions
\begin{gather}
    \mathcal{I}_A = 0 \ ,
    \label{eq:IA_caseIV}\\
    \mathcal{I}_B = h^z_k \ .
    \label{eq:IB_caseIV}
\end{gather}
Finally, the projections onto the reduced sub-sectors are
\begin{gather}
    A_{\pm} = \sum_{i\neq k} h^x_i X_i \ ,
    \label{eq:Apm_caseIV} \\
    B_{\pm} = \sum_{i \neq k} (h^z_i \pm J_{ik}) Z_i + \sum_{\substack{i,j\neq k\\j<i}} J_{ij} Z_i Z_j \pm \mathcal{I}_B \ .
    \label{eq:Bpm_caseIV}
\end{gather}
Despite having carried all our analysis including the possibility of inhomogeneous $x$-fields $h_i^x$, in many practical settings associated to the QA algorithm all $h_i^x$ are, in fact, equal, as considered in~\eqref{eq:A_TF_normalised}. In this scenario, in order to choose the qubit $k$ whose local field we are going to remove we can once again turn to the look-ahead strategy used in cases II and III. However, if homogeneous couplings are assumed, the evaluation of the cost~\eqref{eq:nu_max_cost} is $O(1)$, and only in the case where it turns out to be the best option among all CAS possibilities would the $N-r$ alternative choices of $k$ need to be analysed.


\subsection{Complexity of the recursive CAS reduction}\label{sec:CAS_reduction_complexity}

We now analyse the time and memory complexity of this specific CAS reduction pipeline in detail. 

Gathering the information of the sections above, the brute-force search for the CAS at a given recursion order $r$ takes time 
\begin{equation}
    \zeta^{\text{bf}}(r) = (N-r) + 2 + \sum_{m=2}^{(N-r)/2} 
    \begin{pmatrix}
        N-r \\
        m
    \end{pmatrix} \ ,
\end{equation}
which is reduced to 
\begin{equation}
    \zeta(r) = (N-r) + 2 + O((N-r)^3)
\end{equation}
when we allow for the relaxation to the submodular cost in Eq.~\eqref{eq:Mb_prime_caseIII}. In addition, in many cases type III CAS can be readily discarded as realistic candidates, since they describe a rather structured situation; for example, in a densely connected graph with rather homogeneous couplings, we can expect a type I CAS to be cheaper. This allows doing faster checks to see if type III CAS looks promising before running max-flow, further speeding up the search. 

Despite the efficiency of the search of the CAS within each block, which even becomes lighter as we increase recursion order $r$, we highlight that the number of blocks grows as $2^r$. The tree-like structure of the recursion enables parallelisation, which can alleviate this problem somewhat, but eventually the exponential memory barrier is reached. 

However, for many applications we are only interested in the low-energy sector along the computation, which will not involve all the possible branches. One may notice that, for every CAS reduction, a separation in the energies of the Hamiltonian projected onto the ``$+$" and ``$-$" sectors is introduced. For example, in a type I reduction the ``$-$" sector is expected to carry lower energies, since the identity contributions $\mathcal{I}_A, \mathcal{I}_B$ are always positive (see~\cref{eq:Apm_case1,eq:Bpm_case1,eq:IA_caseI,eq:IB_caseI}). In contrast, type II and III reductions can be expected to have its lowest-energy pseudo-eigenstate in the ``$+$" sector, since $\mathcal{I}_A = \mathcal{I}_B = 0$, $B_- = B_+$ and $\lambda_{\min}(A_+) = \lambda_{\min}(A_-) - 2h_k^x$ (see~\cref{eq:Apm_case2,eq:Bpm_case2,eq:Apm_case3,eq:Bpm_case3}). Finally, in type IV reductions the situation is more unclear, since we have a modification of the local fields of the Ising Hamiltonian~\eqref{eq:Bpm_caseIV} (whose structure is not trivial, unlike in cases II and III) and an identity contribution~\eqref{eq:IB_caseIV} which potentially compete. Let us note that case IV is generally unlikely to arise in non-trivial problems; assuming terms to be roughly of the same order, the normalisation over the (generally) quadratic terms in $B$ will cause $h_i^z, J_{ij} \sim O(1/N)$, while we have $h_i^x \sim O(1/\sqrt{N})$. Thus, a case IV CAS will only be competitive when the number of couplings is linear in $N$ or when local fields dominate. In this scenario, we may consider that the low-energy eigenstates mostly follow the orientations of the $h_i^z$, such that the lowest-energy pseudo-eigenstate will belong to the ``$-$" sector if $h_z^i(s=1) > 0$ and vice versa.

\begin{figure*}[t]
    \centering
    \begin{tikzpicture}
      \node[anchor=south west, inner sep=0] (img) at (0,0)
        {\includegraphics[width=0.8\linewidth]{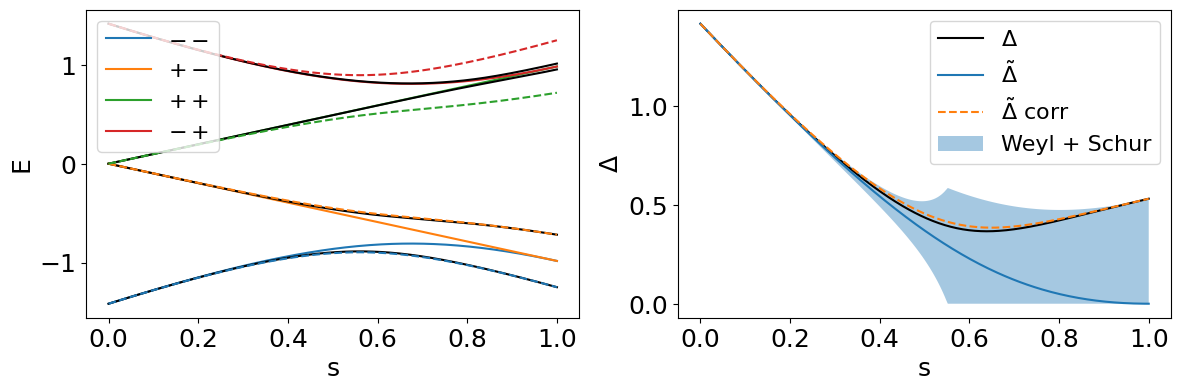}};
      \begin{scope}[x={(img.south east)}, y={(img.north west)}]
        \node[] at (0.03,0.95) {a)};
        \node[] at (0.53,0.95) {b)};
      \end{scope}
    \end{tikzpicture}
    \caption{Example of $N=2$ qubits which have a case II CAS ($A= \frac{-1}{\sqrt{2}}(X_0 + X_1)$, $B=Z_0Z_1 + 0.12Z_0 - 0.15Z_1$). In a), the true energy spectrum is shown in black whereas the computed pseudo-eigenlevels, labelled by their relative paths $\tilde{\xi}$, are shown as coloured, solid lines. The dashed lines correspond to the prediction of modified pseudo-levels considering the interaction between the two participating levels to be maximal, as per Eq.~\eqref{eq:h_eff_corr_pseudo-levels}. b) shows the true gap, the pseudo-gap along with the bounds from Eq.~\eqref{eq:bounds_gap} and the gap prediction from Eq.~\eqref{eq:h_eff_corr_pseudo-levels}.}
    \label{fig:N2_caseII_bounds}
\end{figure*}

Guidance on which will be the block with the lowest energies can be used to build a heuristic that tracks small deviations from the prescribed ``always-lowest-energy" path,
\begin{equation}
    \tilde{\xi}^0 = (\tilde{-}, \tilde{-}, ... \tilde{-})
\end{equation}
where $\tilde{-}$ indicates the lowest-energy block label according to the criteria discussed above. On a deviation of degree 1 we can track paths 
\begin{equation}
    \{\tilde{\xi}^1_{i}\}_{i=0}^{N-2} = \{(\tilde{+}, \tilde{-}, ..., \tilde{-}), (\tilde{-}, \tilde{+}, \tilde{-}, ..., \tilde{-}), ..., 
    (\tilde{-}, ..., \tilde{-}, \tilde{+})\}
\end{equation}
This approach can be generalised to degree $K$, such that
\begin{equation}
    [\tilde{\xi}^K_{\eta_0, ..., \eta_{K-1}}]_i = \left\{ \begin{array}{cl}
       \tilde{-}  & \text{if } i \neq \eta_p \quad \forall \ p=0, ..., K-1\\
        \tilde{+} & \text{otherwise}
    \end{array}
    \right.
\end{equation}
Obtaining the $K$-degree set requires following $\begin{pmatrix}
    N \\ K
\end{pmatrix}$ paths, which is $O(N^K)$. There is, however, no guarantee that the full low-energy pseudo-eigenspace will be contained within a given $K$, as the importance of the energy separation between blocks is also relevant. For example, it could happen that a path with three $\tilde{+}$ of case I with small $\mathcal{I}_A, \mathcal{I}_B$ then contains, further down the line, another case I with a very large $\mathcal{I}_A, \mathcal{I}_B$ which wouldn't have been found following other paths with more $\tilde{-}$.

Nonetheless, the CAS framework provides a program for AQC-inspired classical heuristics that aim to find low-energy solutions of Ising models, which are routinely used to encode a wide variety of relevant optimisation problems. 

\section{Illustrative simulations on small system sizes}\label{sec:small_sizes}

In this Appendix we present some numerical results on small system sizes which, despite being of less interest than the larger scales presented in the main text, we consider provide an instructive picture of the pseudo-spectrum.

We begin by illustrating the pseudo-levels and the resulting bounds on the gap with a minimal example of $N=2$ qubits, which we show in Fig.~\ref{fig:N2_caseII_bounds}. Along with the true gap $\Delta$ and the pseudo-gap $\tilde{\Delta}$ we show the gap prediction that results from considering the eigenvalues of the effective model
\begin{gather}
    \hat{h}_{\text{max}}= \begin{pmatrix}
        \mu_{1}(s) & ||\nu_0^1||_2(s) \\
        ||\nu_0^1||_2(s) & \mu_{0}(s)
    \end{pmatrix} \ ,
    \label{eq:h_eff_corr_pseudo-levels}
\end{gather}
where $\nu_0^1$ is the off-diagonal block connecting the two pseudo-eigenlevels under scrutiny. Note that $\hat{h}_{\text{max}}$ accounts for the largest possible interaction between pseudo-eigenstates, thus providing a reliable upper bound to the gap as long as $\chi_{0k}, \chi_{1k} \ll 1 \ \forall k > 1$.
For this particular instance, the relevant matrix element between the two lowest pseudo-eigenlevels nearly saturates the upper bound provided by $||\nu||_2$, and thus the gap prediction from the model of Eq.~\eqref{eq:h_eff_corr_pseudo-levels} is quite accurate. Consequently, this is not the case for the two highest pseudo-eigenlevels (since they are connected by the same $\nu$), and thus the prediction from Eq.~\eqref{eq:h_eff_corr_pseudo-levels} is further away from the true spectrum than the pseudo-levels themselves. In short, the CAS is satisfied to a good approximation in the high-energy subspace for this problem, i.e., the projectors onto the corresponding pseudo-eigenspaces approximately commute with the relevant CAS operator.


We exemplify the compactness of the description provided by the pseudo-eigenspaces in Fig.~\ref{fig:overlaps_entropy_N5}, where we show the Shannon entropy $S(\tilde{\Pi}_i; \vec{\Pi})$ of the classical probability distribution of the $\ket{\tilde{\mu_k}}$ over the true eigenbasis,  
\begin{equation}
    S(\tilde{\Pi}_i; \vec{\Pi}) = -\sum_k \Tr{\tilde{\Pi}_i \Pi_k} \cdot \log{\left(\Tr{\tilde{\Pi}_i \Pi_k}\right)} \ ,
\end{equation}
for a random Ising instance of $N=5$.
This quantity provides a compact diagnostic of the localisation of the pseudo-levels over the true eigenbasis.
\begin{figure}
    \centering
    \includegraphics[width=0.9\linewidth]{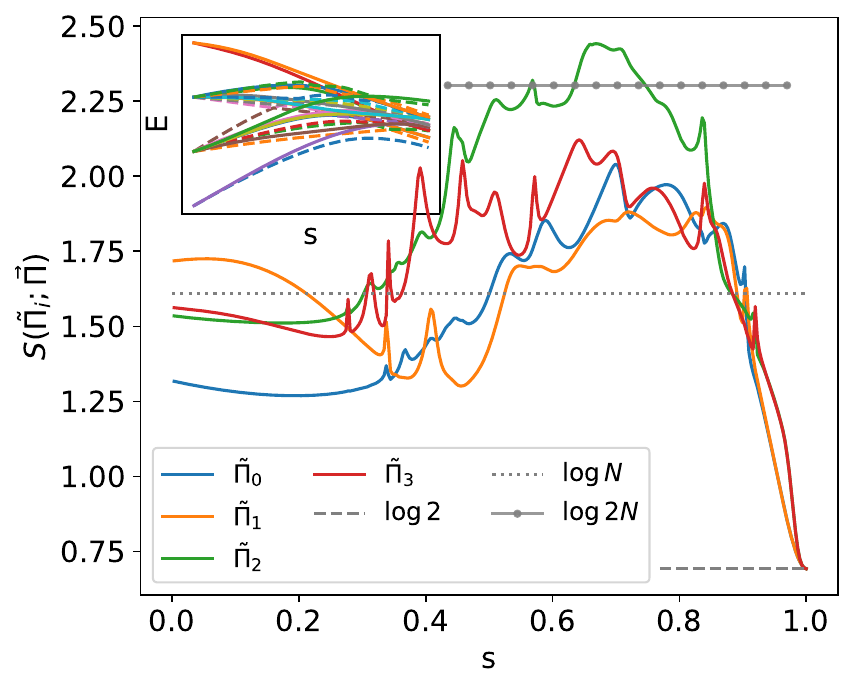}
    \caption{Shannon entropy of the distribution associated with the occupation probabilities of the pseudo-levels 0-3 over the true eigenbasis for a system of $N=5$. The energy spectrum is shown in the inset, where dashed lines refer to the true energies and solid ones to the pseudo-levels. The horizontal lines show the entropy of maximally delocalised vectors with support over 2, $5=N$ and $10=2N$ elements for comparison.}
    \label{fig:overlaps_entropy_N5}
\end{figure}

The particular example of Fig.~\ref{fig:overlaps_entropy_N5} contains a number of case II reductions, which are the source of the final degeneracies of the pseudo-levels under study (note that the entropy is $\log{2}$ at $s=1$). In addition, the entropy of the lowest-lying pseudo-levels remains compatible with a support over the true eigenbasis that is approximately linear in system size. 

\section{Many-body terms in projected sub-blocks}\label{sec:many_body_terms_treatment}

In the reductions of case II and III discussed in Section~\ref{sec:TFIM_example}, a new many-body term arises in the projected subspace, specifically in $A_{\pm}$ (see Eqs.~\eqref{eq:Apm_case2},~\eqref{eq:Apm_case3}). The goal of this Appendix is to discuss the handling of these many-body contributions, which take the projected Hamiltonians outside of the Hamiltonian family originally targeted. 

We will focus the discussion on type II reductions for simplicity; the discussion for case III follows in a similar manner. Without loss of generality, we consider a case II reduction at the first iteration, with the choice $k=N-1$. We will thus study the consequences of the appearance of the highly nonlocal term in the projection of $A$ onto the respective CAS sector, which has the shape
\begin{equation}
    A^{MB}_{\pm} = \pm h_{N-1}^x \prod_{i=0}^{N-2} X_i \ .
    \label{eq:many-body_term}
\end{equation}
First, let us see that the appearance of this many-body term does not change the types of CAS to search for in the projected subspaces according to our term-removal criterion; $\text{DLA}(\Omega_{(A_{\pm}-A^{MB}_{\pm}) \cup B_{\pm}})$ is already full in general, so this term only makes explicit the existence of said Pauli string in the algebra. It may, nonetheless, contribute to the off-diagonal part of the next CAS reduction. 
Consider, for example, that a case IV reduction takes place in the next recursion level. Then, removing a single $X_i$ won't be enough to generate a $\mathbb{Z}_2$ symmetry any more: the many-body term needs to be removed in full as well. In this case, the highly nonlocal term will be absent at the next iteration and the analysis can proceed exactly as described in the main text (and in more detail in Appendix~\ref{sec:CAS_reductions}) again.

Let us now consider the alternative scenario in which the next CAS is of type II or III. Since both of these CAS commute with~\eqref{eq:many-body_term}, no additional cost is induced by its presence on these types of CAS. Moreover, we highlight that, upon a second application of a case II reduction, the many-body term turns into a constant energy shift. To show this, for simplicity we consider the second reduction of type II taking place on $k^\prime = N-2$:
\begin{gather}
    \prod_i^{N-2} X_i (\pm 1) = \prod_i^{N-3} X_i X_{N-2} (\pm 1) \to 
    \nonumber \\
    (\prod_i^{N-3} X_i) (\prod_j^{N-3} X_j) (\pm 1) (\pm 1) \Rightarrow 
    \left\{\begin{array}{cc}
       h^x_k \1  & \text{for - -, + +}  \\
       -h^x_k \1  & \text{for - +, + -}   
    \end{array} \right.
\end{gather}
Note that a new many-body term will arise from the transformation of the local term $X_{k^\prime}$ either way. By the same logic, we can see that the application of a case III will reduce the support of the nonlocal term over the remaining effective degrees of freedom.

In all the previously described scenarios, the many-body term 
goes fully into the off-diagonal part $\nu$ (case IV) or into the diagonal part (cases II and III). If the next reduction involves a type I CAS, however, the many-body term needs to be rotated onto the CAS basis of the case I, generating a block-dependent diagonal component and a block-independent off-diagonal one:
\begin{align}
    U^\dagger_{I} \prod_{i \neq k} X_i X_{k_{I}} U_{I} =& \cos{\varphi_{k_{I}}} \prod_{i \neq k} X_i \tilde{Z}_{k_{I}} - \\
    & - \sin \varphi_{k_{I}} \prod_{i \neq k} X_i \tilde{X}_{k_{I}} \ ,
    \label{eq:caseII_followed_by_caseI}
\end{align}
where $k_I$ denotes the index of the degree of freedom that was chosen for the application of case I and the coefficients are analogous to those in Eq~\eqref{eq:cos_sin_varphi} of the main text. More specifically, we recall that they adopt the shape
\begin{gather}
    \cos{\varphi_{k_{I}}} = \frac{(1-s)h^x_{k_{I}}}{\sqrt{((1-s)h^x_{k_{I}})^2 + (s h^z_{k_{I}})^2}} \ ,\\
    \sin{\varphi_{k_{I}}} =  \frac{sh^z_{k_{I}}}{\sqrt{((1-s)h^x_{k_{I}})^2 + (s h^z_{k_{I}})^2}}   \ .
\end{gather}
When projecting onto the sub-blocks, the sine term will disappear (since we project onto the eigenstates of $\tilde{Z}_{k_{I}}$) but the remaining many-body term with the cosine gets carried over to the sub-block projections.

Finally, note that if the effective number of qubits of the block that had a case II symmetry was 3, the first term in Eq.~\eqref{eq:caseII_followed_by_caseI} constitutes a shift in the $h^x_i$ of the effective blocks. We highlight that similar arguments apply for case III reductions in the presence of many-body terms with the appropriate support, and even for case II reductions at the last levels of recursion.

\end{document}